\documentclass[10pt,journal,twoside,onecolumn,romanappendices]{IEEEtran}

\usepackage{psfrag,epsfig,graphics}
\usepackage{amsmath,amssymb,amsthm,multirow}
\usepackage{cite,subfigure,balance}

\usepackage[ruled, linesnumbered, vlined]{algorithm2e}
\usepackage{bm}
\usepackage{bbm}

\usepackage{color}

\definecolor{violet}{rgb}{0.5,0,0.5}

\newcommand{\cb}[1]{{\boldsymbol{#1}}}
\newcommand{\cp}[1]{\ifmmode {\mathcal{#1}}\else ${\mathcal{#1}}$\fi}

\newcommand{\bsigma}{\boldsymbol{\sigma}}

\newcommand{\bA}{\boldsymbol{A}}
\newcommand{\bB}{\boldsymbol{B}}
\newcommand{\bC}{\boldsymbol{C}}

\newcommand{\bG}{\boldsymbol{G}}
\newcommand{\bH}{\boldsymbol{H}}
\newcommand{\bI}{\boldsymbol{I}}
\newcommand{\bK}{\boldsymbol{K}}
\newcommand{\bM}{\boldsymbol{M}}

\newcommand{\bR}{\boldsymbol{R}}

\newcommand{\bU}{\boldsymbol{U}}

\newcommand{\bO}{\boldsymbol{0}}

\newcommand{\bdf}{\boldsymbol{f}}
\newcommand{\bh}{\boldsymbol{h}}

\newcommand{\bp}{\boldsymbol{p}}
\newcommand{\br}{\boldsymbol{r}}

\newcommand{\bu}{\boldsymbol{u}}
\newcommand{\bv}{\boldsymbol{v}}
\newcommand{\bw}{\boldsymbol{w}}

\newcommand{\bq}{\boldsymbol{q}}
\newcommand{\bg}{\boldsymbol{g}}

\newcommand{\bx}{\boldsymbol{x}}

\newcommand{\diag}{\text{diag}}

\newcommand{\col}{\text{col}}
\newcommand{\sgn}{\text{sgn}}
\newcommand{\vect}{\text{vec}}
\newcommand{\trace}{\text{trace}}

\newcommand{\bSig}{\boldsymbol{\Sigma}}
\newcommand{\bGamma}{\boldsymbol{\Gamma}}

\newcommand{\bPi}{\boldsymbol{\Pi}}
\newcommand{\bLambda}{\boldsymbol{\Lambda}}

\newcommand{\bpsi}{\boldsymbol{\psi}}
\newcommand{\bphi}{\boldsymbol{\phi}}
\newcommand{\E}{{\mathbb{E}}}

\usepackage{tikz}

\newtheorem{remark}{Remark}

\begin{document}

\title{Affine Combination of Diffusion Strategies\\ over Networks}
\author{Danqi Jin, \IEEEmembership{Student Member, IEEE}, \; Jie Chen, \IEEEmembership{Senior Member, IEEE}, \; C{\'e}dric Richard, \IEEEmembership{Senior Member, IEEE}, \; Jingdong Chen, \IEEEmembership{Senior Member, IEEE}, \; Ali H. Sayed, \IEEEmembership{Fellow, IEEE}
\thanks{D. Jin, J. Chen and J. Chen are with Centre of Intelligent Acoustics and Immersive Communications at School of Marine Science and Technology, Northwestern Polytechinical University, Xi'an, China (emails: danqijin@mail.nwpu.edu.cn, dr.jie.chen@ieee.org, jingdongchen@ieee.org). C. Richard is with Universit\'e C\^ote d'Azur, CNRS, France (email: cedric.richard@unice.fr). A. H. Sayed is with the School of Engineering, E\'cole Polytechnique F\'ed\'erale de Lausanne (EPFL), Switzerland (email: ali.sayed@epfl.ch). The work of C. Richard was funded in part by ANR under grant ANR-19-CE48-0002.}\vspace{-0.8cm}}

\maketitle

\begin{abstract}
Diffusion adaptation is a powerful strategy for distributed estimation and learning over networks. Motivated by the concept of combining adaptive filters, this work proposes a combination framework that aggregates the operation of multiple diffusion strategies for enhanced performance. By assigning a combination coefficient to each node, and using an adaptation mechanism to minimize the network error, we obtain a combined diffusion strategy that benefits from the best characteristics of all component strategies simultaneously in terms of excess-mean-square error (EMSE). Analyses of the universality are provided to show the superior performance of affine combination scheme and to characterize its behavior in the mean and mean-square sense. Simulation results are presented to demonstrate the effectiveness of the proposed strategies, as well as the accuracy of theoretical findings.
\end{abstract}

\vspace{-3mm}
\section{INTRODUCTION}
\label{sec:intro}
Distributed adaptive algorithms endow networks with the ability to estimate and track unknown parameters  from streaming data in a collaborative manner. Typical existing techniques include consensus, incremental and diffusion strategies~\cite{Braca2008TSP,Nedic2009,Dimakis2010,Srivastava2011,Sayed2013intr,Sayed2014Proc}. Diffusion strategies have been shown to have superior performance in adaptive scenarios where it is necessary to track drifts in the underlying models through constant step-size adaptation \cite{Tu2012}. There are many variants of diffusion schemes, including diffusion LMS and its multitask counterparts \cite{Chen2014multitask,Chen2015diffusion,Chen2017}, diffusion APA \cite{Li2009}, diffusion RLS~\cite{Cattivelli2008}. In this work, we focus on diffusion techniques due to their enhanced adaptation performance and wider stability ranges.

Diffusion algorithms typically consist of an adaptation step and a combination step. In the adaptation step, each agent updates its iterate by using a local gradient approximation. In the combination step, each agent collects intermediate estimates from its neighbors and fuses them with proper weights. The proper selection of the fusion weights is important for enhanced performance. However, finding an optimal setting for these weights is generally non-trivial.  Several empirical strategies with fixed coefficients have been proposed, including averaging rule, Metropolis rule and relative-degree rule\cite{Sayed2013intr}. Adaptive strategies have also been derived by considering noise levels across the network~\cite{Zhao2012Clustering}, or relationships among agents~\cite{Chen2015diffusion}. Given that different strategies tend to deliver varying performance levels under different operating conditions, it is worth examining the possibility of combining strategies to extract the best performance possible. For example, it is useful to examine whether it is possible to combine two distributed strategies and obtain a new strategy whose performance is superior to its individual components. The question was answered in the affirmative for stand-alone adaptive filters in \cite{Arenas-Garcia2006Mean,Arenas2016Combinations,Kozat2010}. It is more challenging in the context of adaptive networks with a multitude of interacting agents over a graph. We will show nevertheless that this is still possible.

Combination strategies have been successfully used for classical adaptive filters \cite{Arenas-Garcia2006Mean,Arenas-Garcia2006}, multi-kernel learning~\cite{Alain2008SimpleMKL}, as well as modern deep neural network structures~\cite{Szegedy2015}. It is shown in some of these works that using convex combinations \cite{Arenas-Garcia2006,Arenas-Garcia2006Mean} or affine combinations \cite{Bershad2008} of adaptive filters with diversity can lead to filters that combine the advantages of all component filters. Generally, such combination schemes are used to facilitate the selection of filter parameters, to increase robustness against an unknown environment, or to possibly enhance performance beyond the range of each component~\cite{Arenas2016Combinations}. In this paper, we propose affine combination schemes for diffusion strategies. Each agent is designed to run several diffusion strategies in parallel, and to combine their estimates to generate the final estimates. Time-varying affine combination coefficients are set  by minimizing the overall squared instantaneous error. Simulation results show that the proposed algorithms endow the networks with a significantly enhanced performance in the learning process. {Some related works can be found in \cite{Jesus2015} and our previous work \cite{Jin2018ConvexCombination}.} In \cite{Jesus2015}, the authors use a useful convex combination scheme for combining two specific fusion strategies albeit without motivating it or formulating a driving optimization problem. Though similar to \cite{Jesus2015}, in \cite{Jin2018ConvexCombination} a convex combination scheme is proposed for two diffusion strategies. However, both works do not examine the theoretical underpinnings of the algorithms. We should also note that the work proposed here is different from the useful formulation in \cite{Fernandez2017}. This last reference introduces a diffusion scheme for networks with heterogeneous nodes, namely, nodes implementing different adaptive rules or differing in other aspects such as filter structure, length or step-size. In comparison, our work proposes affine combination schemes to combine different diffusion strategies, which can be either homogeneous or heterogeneous. In that sense, the results presented here are more general than \cite{Fernandez2017}.

The main contributions of this work are summarized as follows:
\begin{itemize}
       \item  We introduce a framework for the affine combination of two component diffusion algorithms over networks, and propose two methods for adjusting the combination weights. The framework can be extended to multiple component algorithms as explained in Appendix A.
       \item  We establish the universality of the combined strategy, in a manner that extends a prior universality analysis for the combination of adaptive filters.
       \item We conduct a theoretical analysis of the performance of combined diffusion LMS strategies under some typical simplifying assumptions and approximations. Although the assumptions are not accurate in general, they are nevertheless typical in the context of studying adaptive systems and tend to lead to performance results that match well with practice for sufficiently small step-sizes with white measurement noise and white regressors~\cite{Sayed2008}.
\end{itemize}

\noindent\textbf{Notation.} Normal font $x$ denotes scalars.  Boldface small letters $\bx$ and capital letters $\cb{X}$ denote column vectors and matrices, respectively. The superscript ${(\cdot)}^\top$ denotes the transpose operator. The inverse of a square matrix is denoted by ${(\cdot)}^{-1}$. The mathematical expectation is denoted by $\E\{\cdot\}$. The Gaussian distribution with mean $\mu$ and variance $\sigma^2$ is denoted by $\cp{N}(\mu,\sigma^2)$. The operators $\min\{\cdot\}$ and $\max\{\cdot\}$ return the minimal or maximal value of their arguments. The operator $\diag\{\cdot\}$ extracts the diagonal elements of its matrix argument, or generates a diagonal matrix from its vector argument. The operator $|\cdot|$ returns the absolute value of its  argument. $\bI_{N}$ and $\bO_{N}$ denote identity matrix and zero matrix of size $N\times N$, respectively. All-one vector of length $N$ is denoted by $\mathbbm{1}_N$. Symbol ${\cal C}_i$ stands for cluster $i$, i.e., index set of nodes in the $i$-th cluster. ${\cal{N}}_k$ denotes the neighbors of node $k$, including $k$.

\vspace{0mm}
\section{NETWORK MODEL AND DIFFUSION LMS}
\label{sec:model}
\vspace{-0mm}
\subsection{Network Model}
Consider a connected network consisting of $N$ agents. The problem is to estimate an unknown parameter vector $\bw_k^\star$ of length $L\times 1$ at each agent $k$. Agent $k$ has access to temporal measurement sequences $\{d_{k,n}, \bx_{k,n}\}$, where $d_{k,n}$ denotes a reference signal, and $\bx_{k,n}$ is an $L\times 1$ regression vector with positive-definite covariance matrix. The data at agent $k$ and time instant $n$ are driven by the linear model:
\begin{equation}
	\label{eq:model}
	d_{k,n} = \bx^\top_{k,n}\bw_k^\star + z_{k,n},
\end{equation}
where $z_{k,n}$ is an additive noise. For ease of derivation, we introduce the following assumption \textbf{A1}\footnote{In this paper, we adopt the acronym ``\textbf{A}'' for ``assumption''.} for $z_{k,n}$. Although rarely true in practice, \textbf{A1} simplifies the derivation of algorithm and theoretical analysis, thus it is widely adopted in the context of online learning and adaptation \cite{Sayed2013intr,Sayed2014Proc}.

\textbf{A1}: The additive noise $z_{k,n}$ is zero-mean, stationary, independent and identically distributed (i.i.d.) with variance $\sigma_{z,k}^2$, and independent of any other signal.

To determine the unknown parameter vector $\bw_k^\star$, we consider the following mean-square-error (MSE) cost at agent $k$:
\begin{equation}
\label{eq:cost}
J_k(\bw) = \E\big\{|d_{k,n} - \bx^\top_{k,n}\bw|^2\big\}.
\end{equation}
Observe in \eqref{eq:model} that $J_k(\bw)$ is minimized at $\bw_k^\star$. For single-task problems, each agent in the network estimates the same parameter vector, while for multi-task problems, agents may estimate distinct parameter vectors \cite{Chen2014multitask}.

\vspace{-2mm}
\subsection{Diffusion LMS Algorithm}\label{subsec:DiffusionLMS}

The diffusion LMS algorithm is derived to minimize the following aggregate cost function:
\begin{equation}
	\label{eq:Jglob}
	J^{\text{glob}}(\bw) = \sum_{k=1}^N J_k(\bw)
\end{equation}
in a cooperative manner. The general structure of the diffusion LMS algorithm consists of the following steps:
\begin{align}
    {\bphi}_{k,n} &= \sum_{\ell\in {\cal{N}}_k}a_{1, \ell k}\,{\bw}_{\ell,n}, \label{eq:SystemModel1} \\
	{\bpsi}_{k,n+1} &= {\bphi}_{k,n} +\mu_k\,\sum_{\ell\in {\cal{N}}_k}c_{\ell k}\,\bx_{\ell,n}\,(d_{\ell,n} - \bx^\top_{\ell,n}\,{\bphi}_{k,n}), \label{eq:SystemModel2}  \\
	\bw_{k,n+1} &=\sum_{\ell\in {\cal{N}}_k}a_{2, \ell k}\,{\bpsi}_{\ell,n+1}, \label{eq:SystemModel3}
\end{align}
where $\bw_{k,n+1}$, ${\bpsi}_{k,n+1}$ and ${\bphi}_{k,n}$ are estimates for the unknown parameter vectors $\bw_k^\star$ obtained at different operating stages of the diffusion algorithm, with subscripts $k$ and $n$ denoting node index and time index, respectively, and $\mu_k > 0$ is the step-size at node $k$, the nonnegative coefficients $a_{1,\ell k}, a_{2,\ell k}$ and $c_{\ell k}$ are the $(\ell, k)$-th entries of two left stochastic matrices $\bA_1, \bA_2$ and a right stochastic matrix $\bC$, respectively, satisfying:
\begin{align}
    &\bA_1^\top\mathbbm{1}_N =\mathbbm{1}_N,\, \bA_2^\top\mathbbm{1}_N =\mathbbm{1}_N, \, \bC\mathbbm{1}_N =\mathbbm{1}_N,\label{eq:Constraints1}\\
    & a_{1,\ell k}=0,\;\; a_{2,\ell k}=0,\;\; c_{\ell k}=0\;\; {\rm{if}}\;\; \ell\notin {\cal{N}}_k.\label{eq:Constraints2}
\end{align}
Several adaptive strategies can be obtained as special cases of~\eqref{eq:SystemModel1} to~\eqref{eq:SystemModel3}. Two popular strategies, namely, adapt-then-combine (ATC) and combine-then-adapt (CTA), can be achieved by setting $\bA_1=\bI_N$ and $\bA_2=\bI_N$ respectively.

\vspace{-0mm}
\section{GENERAL COMBINATION FRAMEWORK}
\label{sec:adaptivecombination}

\subsection{Combination framework}
\begin{figure}[!tb]
\begin{minipage}[b]{1.0\linewidth}
  \centering
  \centerline{\includegraphics[width=8cm]{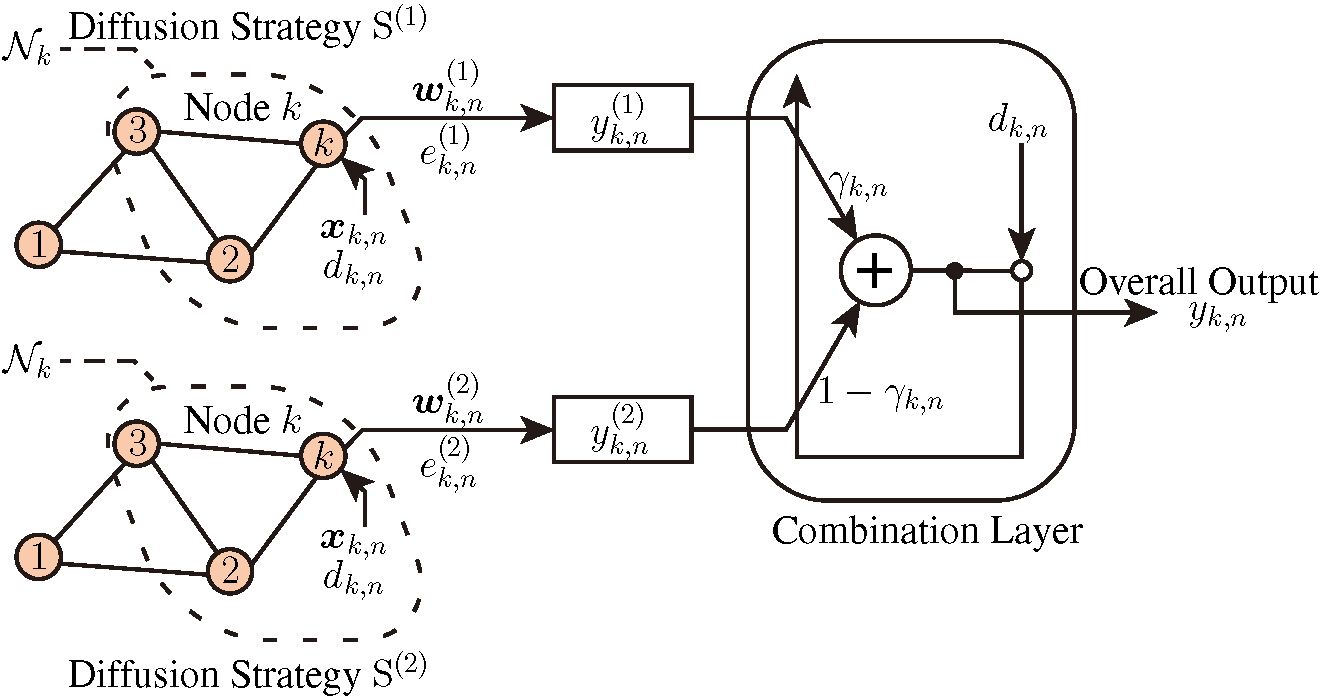}}
  \vspace{-2mm}
\end{minipage}
\caption{Illustration of the combination framework for two diffusion strategies.}
\label{fig1}
\vspace{-5mm}
\end{figure}

Combining two diffusion strategies can be performed with two concurrent adaptive layers: a diffusion strategy layer and a combination layer. The diffusion strategy layer consists of a distributed network running two distinct diffusion strategies, say ${\rm S}^{(1)}$ and ${\rm S}^{(2)}$, individually and simultaneously, with associated quantities denoted by $\bA_1^{(i)}, \bA_2^{(i)}, \bC^{(i)}, \mu^{(i)}$, where the superscript $^{(i)}$ denotes the $i$-th component diffusion strategy ${\rm S}^{(i)}$ with $i =1, 2$. The scheme is illustrated in Fig.~\ref{fig1}. In ${\rm S}^{(1)}$ and ${\rm S}^{(2)}$, the agents have access to identical input and reference signals, and produce each an individual estimate of the optimal weight vector. We associate the combination coefficients $\gamma_{k,n}$ and $1-\gamma_{k,n}$, to ${\rm S}^{(1)}$ and ${\rm S}^{(2)}$, respectively, at each agent $k$ and time instant $n$. The goal of the combination layer is to learn which diffusion strategy performs better at each time instant, by adjusting $\gamma_{k,n}$ in order to optimize the overall network performance.

For each ${\rm S}^{(i)}$, we define the filter output $y_{k,n}^{(i)}$, the estimation error $e_{k,n}^{(i)}$, and the a priori output estimation error $\widetilde{e}_{k,n}^{(i)}$:
\begin{align}
    y_{k,n}^{(i)} &\triangleq \bx_{k,n}^{\top}\,{\bw}_{k,n}^{(i)}, \label{eq:yki}\\
    e_{k,n}^{(i)} &\triangleq d_{k,n} - \bx_{k,n}^{\top}\,{\bw}_{k,n}^{(i)}, \label{eq:eki}\\
    \widetilde{e}_{k,n}^{(i)} &\triangleq \bx_{k,n}^{\top}\,(\bw_k^\star - {\bw}_{k,n}^{(i)}).\label{eq:eaki}
\end{align}
By combining the estimates of two diffusion strategies ${\rm S}^{(1)}$ and ${\rm S}^{(2)}$ at each agent $k$ with coefficients $\gamma_{k,n}$ and $1-\gamma_{k,n}$, we arrive at:
\begin{align}
\bw_{k,n} & \triangleq \gamma_{k,n} \bw_{k,n}^{(1)} + (1-\gamma_{k,n}) \bw_{k,n}^{(2)},\label{eq:wk}\\
y_{k,n} & = \gamma_{k,n} y_{k,n}^{(1)} + (1-\gamma_{k,n}) y_{k,n}^{(2)},\label{eq:yk}\\
e_{k,n} & = \gamma_{k,n} e_{k,n}^{(1)} + (1-\gamma_{k,n}) e_{k,n}^{(2)},\label{eq:ek}\\
\widetilde{e}_{k,n} & = \gamma_{k,n} \widetilde{e}_{k,n}^{(1)} + (1-\gamma_{k,n}) \widetilde{e}_{k,n}^{(2)}\label{eq:eak}.
\end{align}

The problem then becomes one of deriving a strategy for adjusting $\gamma_{k,n}$ based on the minimum mean-square-error (MMSE) criterion. Convex combination schemes require that $\gamma_{k,n}\in[0,1]$. There is no such constraint in affine combination schemes. It is sufficient in this work to consider affine combination schemes to convey the main ideas.

\vspace{-3mm}
\subsection{Affine combination schemes}
\label{sec:affcombination}

The MSE at time instant $n$ of the entire network at the output of the combination layer is defined by:
\begin{equation}
    \label{eq:netMSE}
J^{\rm{MSE}}_n = {1\over 2}\sum_{k=1}^N \E\bigl\{e_{k,n}^2\bigr\}.
\end{equation}
We suggest to adjust $\gamma_{k,n}$ by minimizing \eqref{eq:netMSE}. Using \eqref{eq:ek}, setting the derivative of $J^{\rm{MSE}}_n$ with respect to $\gamma_{k,n}$ to zero, and using the relations:
\begin{align}
&e_{k,n}^{(i)} = \widetilde{e}_{k,n}^{(i)} + z_{k,n}\label{eq:relation1}\\
&\E\{z_{k,n}\} = 0,\label{eq:relation2}
\end{align}
we obtain the optimal value of $\gamma_{k,n}$ in the MMSE sense:
\begin{align}
\hspace{-2mm} \gamma_{k,n}^{\star} \!=\! \frac{\E\bigl\{{(\widetilde{e}_{k,n}^{(2)})}^2\bigr\} \!-\! \E\bigl\{\widetilde{e}_{k,n}^{(1)}\widetilde{e}_{k,n}^{(2)}\bigr\} }{ \E\bigl\{{(\widetilde{e}_{k,n}^{(1)})}^2\bigr\} \!+\! \E\bigl\{{(\widetilde{e}_{k,n}^{(2)})}^2\bigr\} \!-\! 2\E\bigl\{\widetilde{e}_{k,n}^{(1)}\widetilde{e}_{k,n}^{(2)}\bigr\} }.\label{eq:affOptimalSolution}
\end{align}
However, it is not possible to evaluate $\gamma_{k,n}^{\star}$ with~\eqref{eq:affOptimalSolution}  as it requires knowledge of the second-order moments of $\widetilde{e}_{k,n}^{(i)}$. We address this problem by introducing adaptive strategies.

\subsubsection{Affine power-normalized scheme}
Using the gradient descent method to minimize \eqref{eq:netMSE}, and approximating the expectation terms with their instantaneous values, yield the following affine power-normalized LMS iteration \cite{Azpicueta-Ruiz2008}:
\begin{align}
	\gamma_{k,n+1} & = \gamma_{k,n} - {\nu_{\gamma_k}\over \varepsilon + p_{k,n}}{\partial J^{\rm{MSE}}_n\over \partial \gamma_{k,n}}\notag\\
             &\approx \gamma_{k,n} + {\nu_{\gamma_k}\over \varepsilon + p_{k,n}}\,e_{k,n}     \,\bx_{k,n}^{\top}\bigl({\bw}_{k,n}^{(1)} - {\bw}_{k,n}^{(2)}\bigl),\label{eq:affadapta}
\end{align}
where $\varepsilon$ is a small positive parameter, $\nu_{\gamma_k}$ is a positive step-size, and $p_{k,n}$ is a low-pass filtered estimate of the power of $\bx_{k,n}^{\top}\bigl({\bw}_{k,n}^{(1)} - {\bw}_{k,n}^{(2)}\bigl)$ given by:
\begin{equation}
	\label{eq:p(n)}
	p_{k,n} = \eta\,p_{k,n-1} + (1 - \eta){\bigl[\,\bx_{k,n}^{\top}({\bw}_{k,n}^{(1)} - {\bw}_{k,n}^{(2)})\,\bigr]}^2,
\end{equation}
with $0<\eta<1$ a temporal smoothing factor.

\subsubsection{Affine sign-regressor scheme}
Alternatively, we can adopt another normalization scheme for the step-size, leading to the affine sign-regressor LMS iteration \cite{Candido2010}:
\begin{align}
	\gamma_{k,n+1} & = \gamma_{k,n} - {\nu_{\gamma_k}^{\prime}}{\partial J^{\rm{MSE}}_n\over \partial \gamma_{k,n}}\notag\\
             &\approx\gamma_{k,n} + {\nu_{\gamma_k}} e_{k,n}\sgn\bigl\{\bx_{k,n}^{\top}\bigl({\bw}_{k,n}^{(1)} - {\bw}_{k,n}^{(2)}\bigl)\bigl\}\label{eq:adaptaAFFSRLMS}
\end{align}
where $\sgn\{x\}$ is the sign function and $\nu_{\gamma_k}^\prime \!\!=\!\! \frac{\nu_{\gamma_k}}{|\bx_{k,n}^{\top}({\bw}_{k,n}^{(1)} - {\bw}_{k,n}^{(2)})|}$.

\section{THEORETICAL ANALYSIS OF POWER-NORMALIZED SCHEME}\label{sec:theoreticalAff}

\subsection{Universality at steady state}\label{subsec:affuniversality}
We first illustrate the universality of the power-normalized scheme at steady state. In other words, we show that the algorithm results in a combined strategy tracks the best performance of each component strategy.

We start the analysis by defining several quantities to be used later. The EMSE at the output of the combination layer, and the EMSE of each component strategy, at node $k$ and time instant $n$ are defined as:
\begin{align}
	J_{\text{ex},k,n} &\triangleq \E\{{(\widetilde{e}_{k,n})}^2\}	\notag\\
&= \E\bigl\{{\bigl[\gamma_{k,n} \widetilde{e}_{k,n}^{(1)} + (1-\gamma_{k,n}) \widetilde{e}_{k,n}^{(2)}\bigr]}^2\bigr\}\label{eq:EMSE},\\
	J_{\text{ex},k,n}^{(i)} &\triangleq \E\{{(\widetilde{e}_{k,n}^{(i)})}^2\}.\label{eq:comEMSE}
\end{align}
Correspondingly, the EMSE of the whole network, at the output of the combination layer and for each component strategy, are defined by:
\begin{align}
	J_{{\rm ex,net},n} &\triangleq \sum_{k=1}^NJ_{{\rm ex},k,n}\label{eq:EMSEnet}\\
	J_{{\rm ex,net},n}^{(i)} &\triangleq \sum_{k=1}^NJ_{{\rm ex},k,n}^{(i)}\label{eq:EMSEcomponentnet}
\end{align}
respectively. By taking the limit as $n\to\infty$, we obtain the corresponding values at steady-state: $J_{\text{ex},k,\infty}, J_{\text{ex},k,\infty}^{(i)}$, $J_{{\rm ex,net},\infty}$ and $J_{{\rm ex,net},\infty}^{(i)}$. In addition, by resorting to \eqref{eq:eaki} and \eqref{eq:eak}, we have:
\begin{align}
	\bx_{k,n}^{\top}\bigl({\bw}_{k,n}^{(1)} - {\bw}_{k,n}^{(2)}\bigr)
&= \widetilde{e}_{k,n}^{(2)} - \widetilde{e}_{k,n}^{(1)}\label{eq:eakndiffer}
\end{align}
and
\begin{align}
e_{k,n} &= \gamma_{k,n} \widetilde{e}_{k,n}^{(1)}	+ (1-\gamma_{k,n}) \widetilde{e}_{k,n}^{(2)} + z_{k,n}.\label{eq:equivalentekn}
\end{align}
Now we introduce some approximations to be used later to simplify the theoretical analysis:

\textbf{Ap}$_1$\footnote{In this paper, we adopt the acronym ``\textbf{Ap}'' for ``approximation''.}: At steady state, the combination coefficient $\gamma_{k,n}$ is statistically independent of $\widetilde{e}_{k,n}^{(i)}$ and $p_{k,n}$.

\textbf{Ap}$_2$: For a sufficiently large temporal smoothing factor $\eta$, $p_{k,n}$ is statistically independent of $\bx_{k,n}^{\top}{\bw}_{k,n}^{(i)}$, that is, of $\widetilde{e}_{k,n}^{(i)}$.

As indicated in \cite{Arenas-Garcia2006Mean}, approximation \textbf{Ap}$_1$ is reasonable when adopting a decaying step-size $\nu_{\gamma_k}$, and \textbf{Ap}$_2$ is justified when using a large temporal smoothing factor $\eta$. With approximations \textbf{Ap}$_1$ and \textbf{Ap}$_2$, we obtain the following results.

\medskip

\noindent\textbf{Universality Analysis Result 1:} Assume data model \eqref{eq:model}, assumption \textbf{A1} and approximations \textbf{Ap}$_1$, \textbf{Ap}$_2$ hold. Then for any initial conditions with step-size $\nu_{\gamma_k}$ ensuring the stability of power-normalized scheme, the distributed diffusion network with \eqref{eq:affadapta} is universal at steady state, which means that the EMSE of the diffusion network after combination cannot be worse than that of the best component strategies, with
\begin{align}
	J_{{\rm ex,net},\infty} &\leq\min\,\bigl\{J_{{\rm ex,net},\infty}^{(1)}, J_{{\rm ex,net},\infty}^{(2)}\bigr\}.\label{eq:EMSEnetconclusion}
\end{align}

\emph{Proof:} See Appendix B. \hfill $\blacksquare$

\subsection{Mean weight and mean-square behaviors analyses}\label{subsec:MeanandMeanSquareAFF}

We shall now examine the mean and mean-square error behavior of the power-normalized scheme on the basis of the two diffusion LMS strategies referred to in Section \ref{subsec:DiffusionLMS}. Collecting the quantities from the network into block vectors, we have:
\begin{align}
    \bw^\star &\triangleq \col\{\bw_1^\star,\cdots,\bw_N^\star\},\label{eq:blockoptimeanweight}\\
	\bw_n &\triangleq \col\{\bw_{1,n},\cdots,\bw_{N,n}\},\label{eq:blockestimationaftercombine}\\
    \bw_n^{(i)} &\triangleq \col\{\bw_{1,n}^{(i)},\cdots,\bw_{N,n}^{(i)}\},\label{eq:blockweightofcom}
\end{align}
where $\bw^\star$ is the block optimum weight vector, $\bw_n$ and $\bw_n^{(i)}$ are the block weight estimates of the combination layer and component diffusion strategies ${\rm S}^{(i)}$, respectively. Using \eqref{eq:blockestimationaftercombine}, \eqref{eq:blockweightofcom}, and from \eqref{eq:wk} we arrive at:
\begin{equation}
	\label{eq:bwvector}
	\bw_n = \bGamma_n \bw_n^{(1)} + (\bI_{NL} - \bGamma_n) \bw_n^{(2)},
\end{equation}
where $\bGamma_n$ is a diagonal weighting matrix defined by
\begin{equation}
	\label{eq:commatrix}
	\bGamma_n \triangleq \diag\{\gamma_{1,n},\cdots,\gamma_{N,n}\}\otimes \bI_L,
\end{equation}
with symbol $\otimes$ denoting Kronecker product. The weight error vectors at node $k$ for the component diffusion strategies ${\rm S}^{(i)}$ and for the combination layer, as well as those of the entire network, are defined by:
\begin{align}
	\bv_{k,n}^{(i)} &\triangleq \bw_{k,n}^{(i)}-\bw_{k}^{\star},\label{eq:weighterrori}\\
	\bv_{k,n} &\triangleq \bw_{k,n}-\bw_{k}^{\star},\label{eq:weighterrorcombine}\\
	\bv_n^{(i)} &\triangleq \col\{\bv_{1,n}^{(i)},\cdots,\bv_{N,n}^{(i)}\},\label{eq:blockvni}\\
	\bv_n &\triangleq \col\{\bv_{1,n},\cdots,\bv_{N,n}\}.\label{eq:blockvn}
\end{align}
Using \eqref{eq:blockoptimeanweight}--\eqref{eq:blockvn}, vector $\bv_n$ is given by:
\begin{align}
	\bv_{n} = \bw_{n}-\bw^{\star}
    = \bGamma_n\bv_n^{(1)} + (\bI_{NL} - \bGamma_n)\bv_n^{(2)}.\label{eq:bvndecom}
\end{align}
To make the theoretical analysis tractable, we introduce the following assumption \textbf{A2} and approximation \textbf{Ap}$_3$:

\textbf{A2} (\emph{Independent Regressors}): The regression vector $\bx_{k,n}$, generated from a zero-mean random process, is temporally stationary, white (over $n$) and spatially independent (over $k$) with covariance matrix $\bR_{x,k} = \E\{\bx_{k,n}\bx_{k,n}^\top\}>0$.

\textbf{Ap}$_3$: At each time instant $n$, $\gamma_{k,n}$ is statistically independent of $\bw_{k,n}^{(i)}$ for $i = 1, 2$.

Although not true in general, assumption \textbf{A2} is usually adopted to simplify the derivation without constraining the conclusions. Besides, there are several results in the literature showing that performance results obtained under \textbf{A2} match well with actual performance when the step-sizes $\mu_k^{(i)}$ of the component diffusion strategies are sufficiently small \cite{Sayed2008,Sayed2014Adaptation}. \textbf{Ap}$_3$ makes the theoretical analysis tractable. Though actually not true, it does not notably affect the theoretical results, as illustrated by the simulation results.

\subsection*{1) Mean weight behavior analysis}
For the combination layer, taking expectation of \eqref{eq:bvndecom} and using approximation \textbf{Ap}$_3$, we arrive at:
\begin{align}
\E\bigl\{\bv_{n+1}\bigr\}
 \!\!=\!\! \E\bigl\{\bGamma_{n+1}\bigr\}\,\E\bigl\{\bv_{n+1}^{(1)}\bigr\}
 \!+\! \E\bigl\{\bI_{NL} \!-\! \bGamma_{n+1}\bigr\}\,\E\bigl\{\bv_{n+1}^{(2)}\bigr\}.\label{eq:bvnfinal}
\end{align}
We now need to evaluate $\E\bigl\{\bv_{n+1}^{(i)}\bigr\}$. Under assumptions \textbf{A1}-- \textbf{A2}, and following the derivation from \cite{Chen2015diffusion}, we have:
\begin{align}
\bv_{n+1}^{(i)} &= \bB_{n}^{(i)}\bv^{(i)}_n + \bg_n^{(i)}-\br_n^{(i)},\label{eq:compactbv}\\
\E\bigl\{\bv_{n+1}^{(i)}\bigr\}
&= {\overline\bB}^{(i)}\,\E\bigl\{\bv^{(i)}_n\bigr\} - \overline\br^{(i)},\label{eq:compactbvstarexpect2}
\end{align}
with
\begin{align}
\bB_{n}^{(i)}&= {\pmb{\cal A}}_2^{(i)\top}\bigl(\bI_{NL}-\bU^{(i)}\bH^{(i)}_n\bigr)\,{\pmb{\cal A}}_1^{(i)\top},\\
\overline\bB^{(i)} & = {\pmb{\cal A}}_2^{(i)\top}\bigl(\bI_{NL}-\bU^{(i)}\overline\bH^{(i)}\bigr)\,{\pmb{\cal A}}_1^{(i)\top}\label{eq:Bi},\\
{\pmb{\cal A}}_j^{(i)} & = \bA_j^{(i)} \otimes \bI_L,\,\,\forall j=1,2,\\
{\bU}^{(i)} & = \diag\,\bigl\{\mu_1^{(i)},\cdots,\mu_N^{(i)}\bigr\}\otimes\bI_L\label{eq:bU},\\
{\bH}_n^{(i)} & = \diag\,\Bigl\{\sum_{\ell\in {\cal{N}}_k}c_{\ell k}^{(i)}\bx_{\ell,n}\bx^\top_{\ell,n}\Bigr\}_{k =1}^N\label{eq:bH},\\
{\overline\bH}^{(i)}& = \diag\,\Bigl\{\bR_1^{(i)}, \cdots, \bR_N^{(i)}\Bigr\}\label{eq:expbH},\\
\bR_k^{(i)} & \triangleq \sum_{\ell\in {\cal{N}}_k}c_{\ell k}^{(i)}\bR_{x,\ell},\\
\bg_n^{(i)} & = {\pmb{\cal A}}_2^{(i)\top}\bU^{(i)}\bp_{zx,n}^{(i)}\label{eq:gin},\\
\bp_{zx,n}^{(i)}& = \col\,\Bigl\{\sum_{\ell\in {\cal{N}}_k}c_{\ell k}^{(i)}\bx_{\ell,n}z_{\ell,n}\Bigr\}_{k =1}^N \label{eq:pzx},\\
\bh_{u,n}^{(i)} & =\col\,\Bigl\{\sum_{\ell\in {\cal{N}}_k}c_{\ell k}^{(i)}\bx_{\ell,n}\bx^\top_{\ell,n} \bigl(\bw_k^\star-\bw_{\ell}^\star\bigr)\Bigr\}_{k =1}^N,\label{eq:hu}
\end{align}
\begin{align}
& \br_n^{(i)} \triangleq \underbrace{{\pmb{\cal A}}_2^{(i)\top}\bU^{(i)}\bh_{u,n}^{(i)}}_{\br_{\bu,n}^{(i)}}-\notag\\
& \underbrace{\Bigl[{\pmb{\cal A}}_2^{(i)\top}\!\bigl(\bI_{NL}\!-\!\bU^{(i)}\bH^{(i)}_n\bigr)\,\bigl({\pmb{\cal A}}_1^{(i)\top}\!-\!\bI_{NL}\bigr) \!+\! ({\pmb{\cal A}}_2^{(i)\top}\!-\!\bI_{NL})\Bigr]\bw^\star}_{\br_{\bw,n}^{(i)}}\label{eq:brni}\\
&\overline\br^{(i)} \triangleq \E\{\br_n^{(i)}\} = \overline\br_{\bu}^{(i)} - \overline\br_{\bw}^{(i)},
\end{align}
where the symbols with an overhead bar denote the expectation of the corresponding quantities with subscript $n$.

\medskip

\noindent{\textbf{Stability Analysis Result 1:} (Stability in the mean)} Assume data model \eqref{eq:model}, \textbf{A1}--\textbf{A2} and \textbf{Ap}$_3$ hold. Then, for any initial condition, the distributed network with power-normalized diffusion scheme \eqref{eq:affadapta} asymptotically converges in the mean if the step-sizes are chosen to satisfy:
\begin{align}
0\!<\!\mu_k^{(i)}\!<\!\frac{2}{\lambda_{\max}\bigl\{\bR_k^{(i)}\bigr\}},\,\, k=1,\cdots,N\,\,\,\text{and}\,\,\, i=1,2,\label{eq:meanconditionAFF}
\end{align}
where $\lambda_{\max}\{\cdot\}$ denotes the largest eigenvalue of its matrix argument, and if
\begin{align}
0<\nu_{\gamma_k}<1-\eta,\label{eq:muConditionAFFPNLMSMean}
\end{align}
where $\eta$ is the temporal smoothing factor used in \eqref{eq:p(n)}. The asymptotic bias is given by:
\begin{align}
\E\bigl\{\bv_{\infty}\bigr\} & = - \bGamma_{\infty}\,{\bigl(\bI_{NL}-\overline\bB^{(1)}\bigr)}^{-1}\overline\br^{(1)}\notag\\
&\quad -(\bI_{NL} - \bGamma_{\infty})\,{\bigl(\bI_{NL}-\overline\bB^{(2)}\bigr)}^{-1}\overline\br^{(2)}.\label{eq:biasAFF}
\end{align}

\emph{Proof:} See Appendix C. \hfill $\blacksquare$

\subsection*{2) Mean-square behavior analysis}
To perform the mean-square behavior analysis, we introduce the following approximation.

\textbf{Ap}$_4$: At each time instant $n$, $\bGamma_{n+1}$ is statistically independent of $\bB_{n}^{(i)}, \bv^{(i)}_n,  \bg_n^{(i)}$ and $\br_n^{(i)}$ in \eqref{eq:compactbv} for $i=1, 2$.

Although this approximation is not true in general, it will make the analysis tractable.

We shall now evaluate the evolution of $\E\bigl\{{\|\bv_{n+1}\|}^2_{\bSig}\bigr\}$ over time, where $\bSig$ is an arbitrary positive semi-definite matrix, and ${\|\bx\|^2_{\bSig}}\triangleq\bx^\top \bSig \bx$. Using \eqref{eq:bvndecom}, we have:
\begin{align}
& \E\bigl\{{\|\bv_{n+1}\|}^2_{\bSig}\bigr\}  = 2\E\bigl\{\bv_{n+1}^{(1)\top}\bGamma_{n+1}\bSig(\bI_{NL} - \bGamma_{n+1})\bv_{n+1}^{(2)}\bigr\} \notag\\
& \!+\!\E\bigl\{{\bigl\|(\bI_{NL} - \bGamma_{n+1})\bv_{n+1}^{(2)}\bigr\|}^2_{\bSig}\bigr\} +\E\bigl\{{\bigl\|\bGamma_{n+1}\bv_{n+1}^{(1)}\bigr\|}^2_{\bSig}\bigr\}.\label{eq:meansquareanalysis}
\end{align}
\noindent Let
\begin{align}
\bSig^{(1)}_{n+1} & \triangleq\E\bigl\{\bGamma_{n+1}^\top\bSig\bGamma_{n+1}\bigr\},\\
\bSig^{(2)}_{n+1} & \triangleq\E\bigl\{{(\bI_{NL}-\bGamma_{n+1})}^\top\,\bSig\,(\bI_{NL}-\bGamma_{n+1})\bigr\},\\
\bsigma^{(i)}_{n+1}& =\vect\bigl\{\bSig^{(i)}_{n+1}\bigr\},
\end{align}
where $\vect\{\cdot\}$ operator stacks the columns of its matrix argument on top of each other. Under \textbf{Ap}$_3$, the last two terms on the RHS of \eqref{eq:meansquareanalysis} can be written in compact form as $\E\Bigl\{{\bigl\|\bv_{n+1}^{(i)}\bigr\|}^2_{\bSig^{(i)}_{n+1}}\Bigr\}$ with $i = 1, 2$. They can be evaluated compactly as:
\begin{align}
& \E\Bigl\{{\bigl\|\bv_{n+1}^{(i)}\bigr\|}^2_{\bsigma^{(i)}_{n+1}}\Bigr\} = \E\Bigl\{{\bigl\|\bv_{n}^{(i)}\bigr\|}^2_{\bK^{(i)}\bsigma^{(i)}_{n+1}}\Bigr\}+\notag\\
&\quad {\bigl[\vect\{\bG^{(i)\top}\} \bigr]}^\top\bsigma^{(i)}_{n+1} +
\bdf\bigl(\overline\br^{(i)},\bSig^{(i)}_{n+1},\E\bigl\{\bv_n^{(i)}\bigr\}\bigr),\label{eq:meansquareapprox}
\end{align}
where ${\|\cdot\|}_{\bSig^{(i)}_{n+1}}^2$ and ${\|\cdot\|}_{\bsigma^{(i)}_{n+1}}^2$ are used interchangeably, with:
\begin{align}
\hspace{-3mm} \bdf(\br_n^{(i)},\bSig^{(i)}_{n+1},\bv_n^{(i)}) & \triangleq {\bigl\|\br_{n}^{(i)}\bigr\|}^2_{\bSig^{(i)}_{n+1}}\!\!-\!\!2\br_n^{(i)\top}\bSig^{(i)}_{n+1}\bB^{(i)}_n\bv_n^{(i)}\label{eq:Bf}\\
\bK^{(i)} & \approx \overline\bB^{(i)\top}\otimes\overline\bB^{(i)\top}\label{eq:ApproxKi}\\
\bG^{(i)} & \triangleq\E\bigl\{\bg_n^{(i)}\bg_n^{(i)\top}\bigr\}\label{eq:BG}.
\end{align}
The derivation of equation \eqref{eq:meansquareapprox} is provided in Appendix D.

For the first term on RHS of \eqref{eq:meansquareanalysis}, since it has a similar structure to \eqref{eq:meansquarecomicompact}, the analysis is omitted. Using \eqref{eq:compactbv}, following the routine \eqref{eq:meansquarecomicompact}--\eqref{eq:Efapprox}, and ignoring second-order terms in the step-size, we finally arrive at:
\begin{align}
\hspace{-2mm}&\E\bigl\{\bv_{n+1}^{(1)\top}\bSig_{\text{x},n+1}\,\bv_{n+1}^{(2)}\bigr\}= \notag\\
\hspace{-2mm}&\quad\E\bigl\{\bv_{n}^{(1)\top}\bSig_{\text{xc},n+1}\bv_{n}^{(2)}\bigr\}+{\bigl[\vect\{\bG_{\text{x}}^\top\}\bigr]}^\top\bsigma_{\text{x},n+1}+\notag\\
\hspace{-2mm}&\quad\bdf_{\text{x}}\bigl(\overline\br^{(1)},\overline\br^{(2)},\bSig_{\text{x},n+1},\E\bigl\{\!\bv_n^{(1)}\!\bigr\},\E\bigl\{\!\bv_n^{(2)}\!\bigr\},\overline\bB^{(1)},\overline\bB^{(2)}\bigr),\label{eq:crosstermiteration}
\end{align}
where
\begin{align}
&\bdf_{\text{x}}\bigl(\overline\br^{(1)},\overline\br^{(2)},\bSig_{\text{x},n+1},\E\bigl\{\bv_n^{(1)}\bigr\},\E\bigl\{\bv_n^{(2)}\bigr\},\overline\bB^{(1)},\overline\bB^{(2)}\bigr)\notag\\
&\triangleq\overline\br^{(1)\top}\bSig_{\text{x},n+1}\,\overline\br^{(2)} -\E\bigl\{\bv_n^{(1)\top}\bigr\}\overline\bB^{(1)\top}\bSig_{\text{x},n+1}\overline\br^{(2)}\notag\\
&\quad-\overline\br^{(1)\top}\bSig_{\text{x},n+1}\overline\bB^{(2)}\E\bigl\{\bv_n^{(2)}\bigr\}\label{eq:bdfcrossdefinition}
\end{align}
with
\begin{align}
\bSig_{\text{x},n+1}&\triangleq\E\bigl\{\bGamma_{n+1}\,\bSig\,(\bI_{NL} - \bGamma_{n+1})\bigr\},\\
\bSig_{\text{xc},n+1}&\triangleq\vect^{-1}\,\bigl\{\bK_{\text{x}}\,\bsigma_{\text{x},n+1}\bigr\},\\
\bsigma_{\text{x},n+1}&\triangleq\vect\,\bigl\{\bSig_{\text{x},n+1}\bigr\},\\
\bK_{\text{x}}& \approx \overline\bB^{(2)\top}\otimes\overline\bB^{(1)\top},\label{eq:ApproxKcross}\\
\bG_{\text{x}}& \triangleq \E\bigl\{\bg_{n}^{(2)}\bg_{n}^{(1)\top}\bigr\},
\end{align}
and $\vect^{-1}\,\{\cdot\}$ is the inverse vectorization operator.

Finally, using \eqref{eq:meansquareapprox} for $i = 1, 2$ and \eqref{eq:crosstermiteration} in \eqref{eq:meansquareanalysis}, we obtain the explicit expression of the weighted mean-square behavior of the power-normalized diffusion scheme.

\medskip

\noindent{\textbf{Stability Analysis Result 2:} (Mean-square Stability)} Assume data model \eqref{eq:model} and \textbf{A1}--\textbf{A2}, \textbf{Ap}$_3$--\textbf{Ap}$_4$ hold. Assume further that the step-sizes $\mu_k^{(i)}$ for $i = 1, 2$ are sufficiently small such that condition \eqref{eq:meanconditionAFF} is satisfied and approximations \eqref{eq:ApproxKi}, \eqref{eq:ApproxKcross}, \eqref{eq:Efapprox} are justified by ignoring higher powers of the step-size. Furthermore, assume that the step-size $\nu_{\gamma_k}$ at the combination layer satisfies the condition:
\begin{align}
0<\nu_{\gamma_k}<\frac{1-\eta}{3}\label{eq:muConditionAFFPNLMSMeanaquarestate}
\end{align}
to ensure mean-square stability of the power-normalized scheme. Then, for any initial conditions, the distributed network with doubly stochastic matrices $\bA_1^{(i)}, \bA_2^{(i)}$, of which both columns and rows add up to one, and scheme \eqref{eq:affadapta} is mean-square stable if $\rho\bigl(\bI_{NL}-\bU^{(i)}\overline\bH^{(i)}\bigr)<1$, which is further guaranteed for sufficiently small step-sizes that also satisfy condition \eqref{eq:meanconditionAFF}.

\emph{Proof:} See Appendix E. \hfill $\blacksquare$

{\remark(Transient MSD):} We shall adopt the mean-square-deviation (MSD) learning curve of the entire network, defined by $\xi_{n+1}\triangleq \E\bigl\{{\|\bv_{n+1}\|}^2_{\frac{1}{N}\bI_{NL}}\bigr\}$, as a metric to evaluate the performance of diffusion networks. We observe that although the weighting matrix $\bSig$ of $\E\bigl\{{\|\bv_{n+1}\|}^2_{\bSig}\bigr\}$ in \eqref{eq:meansquareanalysis} may be constant over time, such as choosing $\bSig = \frac{1}{N}\bI_{NL}$ in evaluating the MSD, matrices $\bSig^{(i)}_{n+1}$ and $\bSig_{\text{x}, n+1}$ become time-variant since $\bGamma_{n+1}$ varies over time. Thus, it is almost impossible to derive a recursion to relate $\E\bigl\{{\|\bv_{n+1}\|}^2_{\bSig}\bigr\}$ and $\E\bigl\{{\|\bv_{n}\|}^2_{\bSig}\bigr\}$ directly. To evaluate $\E\bigl\{{\|\bv_{n+1}\|}^2_{\bSig}\bigr\}$, we must evaluate \eqref{eq:meansquareapprox} and \eqref{eq:crosstermiteration}, in which $\E\bigl\{\bigl\|\bv_{n}^{(i)}\bigr\|^2_{\bK^{(i)}\bsigma^{(i)}_{n+1}}\bigr\}$ and $\E\bigl\{\bv_{n}^{(1)\top}\bSig_{\text{xc},n+1}\bv_{n}^{(2)}\bigr\}$ are calculated iteratively as shown by expressions \eqref{eq:MSDIteration} and \eqref{eq:MSDIterationCross} in Appendix~F, while the remaining terms can be evaluated directly.

{\remark(Steady-state MSD):}For sufficiently small step-sizes satisfying conditions \eqref{eq:meanconditionAFF} and \eqref{eq:muConditionAFFPNLMSMeanaquarestate} to ensure the mean and mean-square stabilities of the power-normalized diffusion scheme, we can obtain the explicit expression of the steady-state MSD as shown by expression \eqref{eq:netSteadyMSD} in Appendix G.

\emph{Proof:} See Appendix G. \hfill $\blacksquare$\hfill

\subsection{Mean and mean-square behaviors of $\gamma_{k,n}$}\label{subsec:combinationlayerAff}
In order to evaluate the mean and mean-square behavior of the power-normalized diffusion scheme, we must determine the mean and mean-square behavior of $\bGamma_{n}$ at the combination layer, which are obtained by evaluating those of $\gamma_{k,n}$ since $\bGamma_{n}$ is diagonal. To make the analysis tractable, we introduce the following approximations.

\textbf{Ap}$_5$: The combination coefficient $\gamma_{k,n}$ varies slowly enough so that the correlation between $\gamma_{k,n}$ and $\widetilde{e}_{k,n}^{(m)}\widetilde{e}_{k,n}^{(n)}$ for $m, n = 1, 2$ can be ignored.

\textbf{Ap}$_6$: For a large enough temporal smoothing factor $\eta$, $p_{k,n}$ is statistically independent of $\gamma_{k,n}$.

\textbf{Ap}$_7$: The a priori errors $\widetilde{e}_{k,n}^{(1)}$ and $\widetilde{e}_{k,n}^{(2)}$ are jointly Gaussian with zero-mean, which implies \cite{papoulis2002}:
\begin{align}
\E\bigl\{{(\widetilde{e}_{k,n}^{(i)})}^4\bigr\} & = 3{\bigl(J_{\text{ex},k,n}^{(i)}\bigr)}^2,\,\,\forall \,\,i=1,2,\\
\E\bigl\{{(\widetilde{e}_{k,n}^{(1)})}^3{(\widetilde{e}_{k,n}^{(2)})}^1\bigr\} & = 3J_{\text{ex},k,n}^{(1)}J_{\text{ex},k,n}^{(1,2)},\\
\E\bigl\{{(\widetilde{e}_{k,n}^{(1)})}^1{(\widetilde{e}_{k,n}^{(2)})}^3\bigr\} & = 3J_{\text{ex},k,n}^{(1,2)}J_{\text{ex},k,n}^{(2)},\\
\E\bigl\{{(\widetilde{e}_{k,n}^{(1)})}^2{(\widetilde{e}_{k,n}^{(2)})}^2\bigr\} & = 2{\bigl(J_{\text{ex},k,n}^{(1,2)}\bigr)}^2 + J_{\text{ex},k,n}^{(1)}J_{\text{ex},k,n}^{(2)}.
\end{align}

Approximation \textbf{Ap}$_5$ is commonly adopted in transient analysis of affine combinations of two adaptive filters \cite{Candido2010}, which also coincides with simulation result that $\gamma_{k,n}$ converges slowly compared to the variations of input signal $\bx_{k,n}$, thus to the variations of a priori errors $\widetilde{e}_{k,n}^{(i)}$. Although not true in general, \textbf{Ap}$_6$ makes the analysis tractable. Although may be violated in general, assumption \textbf{Ap}$_7$ is frequently adopted to facilitate the transient analysis of adaptive filters \cite{Chen2016ZALMS,Sayed2014Proc,haykin2005,Candido2010,Nascimento2009,Silva2010,Jin2018}. It becomes more reasonable for small step-sizes and long filters~\cite{Sayed2014Proc}.

Starting from \eqref{eq:affadapta}, and following the derivation in Appendix~H, we arrive at the following {stability analysis results}.

\medskip

\noindent\textbf{Stability Analysis Result 3: (Stability in the Mean)} Assume data model \eqref{eq:model}, assumption \textbf{A1} and approximations \textbf{Ap}$_2$, \textbf{Ap}$_5$, \textbf{Ap}$_6$ hold. Then for any initial conditions, the power-normalized scheme \eqref{eq:affadapta} asymptotically converges in the mean if the step-sizes ${\nu_{\gamma_k}}$ are chosen to satisfy condition \eqref{eq:muConditionAFFPNLMSMean}. Besides, the mean behavior of $\gamma_{k,n}$ is evaluated as \eqref{eq:meanIterationAFF} in Appendix H, with steady-state value $\E\{\gamma_{k,\infty}\}$ given by equation~\eqref{eq:affStationarySolution}.

\emph{Proof:} See Appendix H. \hfill $\blacksquare$\hfill

\medskip

\noindent\textbf{Stability Analysis Result 4: (Mean-square Stability)} Assume data model \eqref{eq:model}, assumption \textbf{A1} and approximations \textbf{Ap}$_2$, \textbf{Ap}$_5$--\textbf{Ap}$_7$ hold. Then for any initial conditions, the power-normalized scheme \eqref{eq:affadapta} is mean-square stable if the step-sizes ${\nu_{\gamma_k}}$ are chosen to satisfy condition \eqref{eq:muConditionAFFPNLMSMeanaquarestate}. Besides, the transient mean-square behavior of $\gamma_{k,n}$ is evaluated by \eqref{eq:AFFPNLMSExpectation} in Appendix H, with steady-state value $\E\{\gamma_{k,\infty}^2\}$ given by equation~\eqref{eq:gammasquaresteadyAff}.

\begin{figure*}[!t]
\normalsize
\begin{equation}
\E\{\gamma_{k,\infty}^2\} \!\!=\!\! \frac{\bar\nu_{\infty}(J_{\text{ex},k,\infty}^{(2)}\!\!+\!\!\sigma_{z,k}^2)\bigl(\triangle J_{k,\infty}^{(1)}\!\!+\!\!\triangle J_{k,\infty}^{(2)}\bigr) \!\!+\! 2\bar\nu_{\infty}{(\triangle J_{k,\infty}^{(2)})}^2 \!\!+\! 2\E\{\gamma_{k,\infty}\}\bigl[\triangle J_{k,\infty}^{(2)}\!\!-\!\!3\bar\nu_{\infty}\triangle J_{k,\infty}^{(2)}(\triangle J_{k,\infty}^{(1)}\!\!+\!\!\triangle J_{k,\infty}^{(2)})\bigr]}{2\bigl(\triangle J_{k,\infty}^{(1)}+\triangle J_{k,\infty}^{(2)}\bigr)-3\bar\nu_{\infty}\,{\bigl(\triangle J_{k,\infty}^{(1)}+\triangle J_{k,\infty}^{(2)}\bigr)}^2}\label{eq:gammasquaresteadyAff}
\end{equation}
\hrulefill
\end{figure*}

\emph{Proof:} See Appendix H. \hfill $\blacksquare$\hfill

\section{THEORETICAL ANALYSIS OF SIGN-REGRESSOR SCHEME}\label{sec:theoreticalAffSR}
By following the same routine as in Section \ref{sec:theoreticalAff}, we conduct theoretical analysis for the sign-regressor diffusion scheme.

\vspace{-3mm}
\subsection{Universality at steady state}\label{subsec:affSRuniversality}
Taking expectation of \eqref{eq:adaptaAFFSRLMS}, and using \eqref{eq:eakndiffer}, \eqref{eq:equivalentekn}, we obtain:
\begin{align}
	\E\{\gamma_{k,n+1}\}
&=\E\bigl\{\gamma_{k,n}\bigr\} + {\nu_{\gamma_k}}\, \E\bigl\{\widetilde{e}_{k,n}^{(2)}\,\sgn\bigl\{\widetilde{e}_{k,n}^{(2)} - \widetilde{e}_{k,n}^{(1)}\bigr\}\notag\\&\quad-\gamma_{k,n}\bigl|\widetilde{e}_{k,n}^{(2)} - \widetilde{e}_{k,n}^{(1)}\bigr|\bigr\}.\label{eq:adaptExpAffSR}
\end{align}
From \eqref{eq:adaptExpAffSR}, the stationary point of $\gamma_{k,n}$ is reached if
\begin{align}
\E\Bigl\{\!\widetilde{e}_{k,n}^{(2)}\sgn\bigl\{\!\widetilde{e}_{k,n}^{(2)}\!\!\!-\!\widetilde{e}_{k,n}^{(1)}\!\bigr\}\!-\!\gamma_{k,n}\bigl|\widetilde{e}_{k,n}^{(2)}\!\!\!-\! \widetilde{e}_{k,n}^{(1)}\bigr|\!\Bigr\}\!=\!0.\label{eq:optimalconditionaffSR}
\end{align}
Using the independence approximation \textbf{Ap}$_1$, \eqref{eq:optimalconditionaffSR} gives a closed-form solution of $\E\{\gamma_{k,n}\}$ at steady-state as:
\begin{align}
\E\{\gamma_{k,\infty}\} &= \frac{\E\bigl\{\widetilde{e}_{k,\infty}^{(2)}\,\sgn\bigl\{\widetilde{e}_{k,\infty}^{(2)} - \widetilde{e}_{k,\infty}^{(1)}\bigr\}\bigr\}}{\E\bigl\{\bigl|\widetilde{e}_{k,\infty}^{(2)} - \widetilde{e}_{k,\infty}^{(1)}\bigr|\bigr\}}.\label{eq:affSRStationarySolution}
\end{align}
We further resort to the joint Gaussian approximation \textbf{Ap}$_7$, which leads to approximations \eqref{eq:approximationresult1} and \eqref{eq:approximationresult2} further ahead in Section \ref{subsec:combinationlayerAffSR}. Using \eqref{eq:approximationresult1} and \eqref{eq:approximationresult2}, \eqref{eq:affSRStationarySolution} simplifies to \eqref{eq:affStationarySolution}. With \eqref{eq:affStationarySolution}, the proof of universality for the sign-regressor diffusion scheme at steady state is identical to that for power-normalized diffusion scheme.

\vspace{-3mm}
\subsection{Mean weight and mean-square behaviors analyses}\label{subsec:MeanandMeanSquareAffSR}
Since there is almost no difference in the mean weight and mean-square behaviors of the sign-regressor and power-normalized schemes, we only provide the main conclusions here. Under assumptions \textbf{A1}--\textbf{A2}, approximations \textbf{Ap}$_3$--\textbf{Ap}$_4$ and following the same routine as that in Section \ref{subsec:MeanandMeanSquareAFF}, the mean and mean-square behaviors of the weight error vector $\bv_{n+1}$ at combination layer of the sign-regressor diffusion scheme are given by \eqref{eq:bvnfinal} and \eqref{eq:meansquareanalysis}, respectively.

\medskip

\noindent{\textbf{Stability Analysis Result 5:} (Stability in the Mean)} Assume data model \eqref{eq:model}, \textbf{A1}--\textbf{A2} and \textbf{Ap}$_3$ hold. Then for any initial conditions, the distributed network with sign-regressor diffusion scheme \eqref{eq:adaptaAFFSRLMS} asymptotically converges in the mean if the step-sizes of network are chosen to satisfy \eqref{eq:meanconditionAFF}, and if the step-sizes at the combination layer are chosen to satisfy
\begin{align}
0<{\nu_{\gamma_k}}<\sqrt{\frac{\pi}{2\cdot\max\limits_n\bigl\{{\triangle J_{k,n}^{(1)}+\triangle J_{k,n}^{(2)}}\bigr\}}}.\label{eq:AffSRstepsizeConditionMean}
\end{align}
The asymptotic bias is given by \eqref{eq:biasAFF}.

\medskip

\noindent{\textbf{Stability Analysis Result 6:} (Mean-square Stability)} Assume data model \eqref{eq:model} and \textbf{A1}--\textbf{A2}, \textbf{Ap}$_3$--\textbf{Ap}$_4$ hold. Assume further that the step-sizes $\mu_k^{(i)}$ for $i = 1, 2$ are sufficiently small such that condition \eqref{eq:meanconditionAFF} is satisfied and approximations \eqref{eq:ApproxKi}, \eqref{eq:ApproxKcross}, \eqref{eq:Efapprox} hold. Furthermore, assume that the step-sizes $\nu_{\gamma_k}$ at the combination layer satisfy condition
\begin{align}
0<{\nu_{\gamma_k}}<\sqrt{\frac{2}{\pi\max\limits_n\bigl\{{\triangle J_{k,n}^{(1)}+\triangle J_{k,n}^{(2)}}\bigr\}}}\label{eq:AffSRmeansquareConditionMeanaquarestate}
\end{align}
to ensure the mean-square stability of the sign-regressor diffusion scheme. Then for any initial conditions, the distributed network with doubly stochastic matrices $\bA_1^{(i)}, \bA_2^{(i)}$ and scheme \eqref{eq:adaptaAFFSRLMS} is mean-square stable if $\rho\bigl(\bI_{NL}-\bU^{(i)}\overline\bH^{(i)}\bigr)<1$, which is further guaranteed by sufficiently small step-sizes that also satisfy condition \eqref{eq:meanconditionAFF}.

\subsection{Mean and mean-square behaviors of $\gamma_{k,n}$}\label{subsec:combinationlayerAffSR}

Starting from \eqref{eq:adaptaAFFSRLMS}, and following the derivation in Appendix~I, we arrive at the following {stability analysis result}.

\medskip

\noindent\textbf{Stability Analysis Result 7: (Stability in the Mean)} Assume data model \eqref{eq:model} and \textbf{A1}, \textbf{Ap}$_5$, \textbf{Ap}$_7$ hold. Then for any initial conditions, the sign-regressor scheme \eqref{eq:adaptaAFFSRLMS} asymptotically converges in the mean if the step-sizes ${\nu_{\gamma_k}}$ are chosen to satisfy condition \eqref{eq:AffSRstepsizeConditionMean}. Besides, the mean behavior of $\gamma_{k,n}$ is evaluated by \eqref{eq:IterTheoreticalAffSRresult} in Appendix I, with steady-state value $\E\{\gamma_{k,\infty}\}$ given by equation~\eqref{eq:affStationarySolution}.

\emph{Proof:} See Appendix I. \hfill $\blacksquare$\hfill

\medskip

\noindent\textbf{Stability Analysis Result 8: (Mean-square Stability)} Assume data model \eqref{eq:model} and \textbf{A1}, \textbf{Ap}$_5$, \textbf{Ap}$_7$ hold. Then for any initial conditions, the sign-regressor scheme \eqref{eq:adaptaAFFSRLMS} is mean-square stable if the step-sizes ${\nu_{\gamma_k}}$ are chosen to satisfy condition \eqref{eq:AffSRmeansquareConditionMeanaquarestate}. Besides, the transient mean-square behavior of $\gamma_{k,n}$ is evaluated by \eqref{eq:adaptAffSRmeansquareexpect} in Appendix I, with steady-state value $\E\{\gamma_{k,\infty}^2\}$ given by equation~\eqref{eq:AffSRmeansquareSteadyState}.

\begin{figure*}[!t]
\normalsize
\begin{equation}
\E\{\gamma_{k,\infty}^2\}
=\frac{\nu_{\gamma_k}\bigl(J_{\text{ex},k,\infty}^{(2)} + \sigma_{z,k}^2\bigr) + 2\E\{\gamma_{k,\infty}\}\biggl[\triangle J_{k,\infty}^{(2)}\sqrt{\frac{2}{\pi\bigl({\triangle J_{k,\infty}^{(1)}+\triangle J_{k,\infty}^{(2)}}\bigr)}}-\nu_{\gamma_k}\triangle J_{k,\infty}^{(2)}\biggr]}{\sqrt{{8\,\bigl({\triangle J_{k,\infty}^{(1)}+\triangle J_{k,\infty}^{(2)}}\bigr)}/{\pi}}-\nu_{\gamma_k}\bigl({\triangle J_{k,\infty}^{(1)}+\triangle J_{k,\infty}^{(2)}}\bigr)}\label{eq:AffSRmeansquareSteadyState}
\end{equation}
\hrulefill
\end{figure*}

\emph{Proof:} See Appendix I. \hfill $\blacksquare$\hfill

\section{SIMULATION RESULTS}
\label{sec:simulation}

In this section, we present simulation results to illustrate the proposed combination schemes and theoretical results. All simulated curves were averaged over $100$ Monte Carlo runs.

\begin{figure}[!t]
\begin{minipage}[b]{1.0\linewidth}
  \centering
  \centerline{\includegraphics[width=6.5cm]{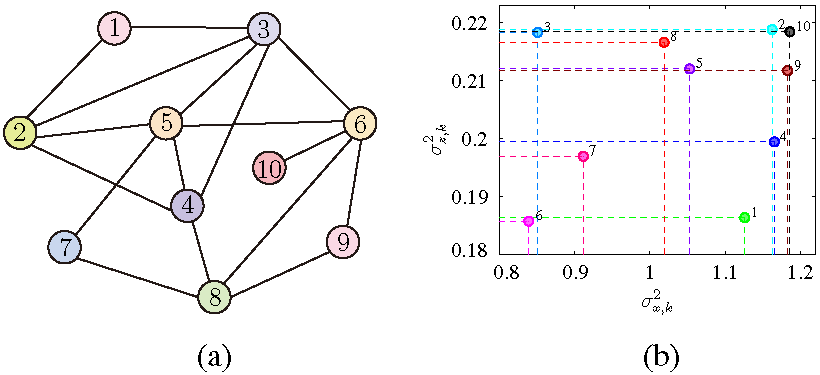}}
\end{minipage}
\caption{Network topology and associated input variances and noise variances. (a) Network topology; (b) Agent input and noise variances.}
\label{fig2}
\end{figure}

\subsection{Affine combination schemes validation}\label{subsec:simulationAff}

Consider a non-stationary system identification scenario with $\bw^\star_k$ varying over time. The distributed network consisted of $N=10$ nodes with connection topology depicted in Fig.~\ref{fig2}(a). The regressors were generated from a multivariate Gaussian distribution with zero-mean and covariance matrix $\bR_{x,k} = \sigma_{x,k}^2\bI_{50}$. The noise signals were generated from Gaussian distribution $\cp{N}(0,\sigma_{z,k}^2)$. Variances $\sigma_{x,k}^2$ and $\sigma_{z,k}^2$ at each agent were generated randomly as depicted in Fig.~\ref{fig2}(b).

\subsubsection{Affine combination of diffusion LMS strategies}\label{subsec:affineValidation}
We first considered two ATC diffusion LMS strategies as component strategy.
Without loss of generality, matrices $\bC$ and $\bA_1^{(i)}$ were set to the identity matrix. As a result, vector $\bphi_{k,n}$ coincides with $\bw_{k,n}$ in equations \eqref{eq:SystemModel1} and \eqref{eq:SystemModel2}. We considered two groups of combination matrices $\bA_2^{(i)}$: static combination matrices with $\bA_2^{(1)}=\bI_N$ and averaging rule \cite{Sayed2013intr} for $\bA_2^{(2)}$, and adaptive combination matrices for $\bA_2^{(1)}, \bA_2^{(2)}$ given in \cite{Chen2015diffusion} and \cite{Zhao2012Clustering}, with $(\ell, k)$-th entries at time instant $n+1$ given by:
\begin{align}
a_{2,\ell k}^{(1)}(n\!+\!1) &\!=\! \frac{\|\bpsi_{k,n+1}+\mu_k\bq_{k,n}-\bpsi_{\ell,n+1}\|^{-2}}{\sum_{j\in{\cal N}_k}\!\|\bpsi_{k,n+1}\!+\!\mu_k\bq_{k,n}\!-\!\bpsi_{j,n+1}\|^{-2}}\label{eq:adaMatric-a21}\\
a_{2,\ell k}^{(2)}(n\!+\!1) &= \frac{\zeta_{\ell k, n+1}^{-2}}{\sum_{j\in{\cal N}_k}\zeta_{j k, n+1}^{-2}}\label{eq:adaMatric-a22}
\end{align}
${\rm for}\,\,\ell\in{\cal N}_k$. We use the index $(n+1)$ in \eqref{eq:adaMatric-a21} and \eqref{eq:adaMatric-a22} to highlight the time-variant nature of adaptive combination matrices, and quantities $\bq_{k,n}$ and $\zeta_{\ell k, n+1}$ are evaluated as:
\begin{align}
\bq_{k,n} &\!=\! [d_{k,n}-\bx_{k,n}^{\top}\bpsi_{k,n+1}]\cdot\bx_{k,n}\\
\zeta_{\ell k, n+1}^2 &= (1-\tau_k)\cdot\zeta_{\ell k, n}^2 + \tau_k\cdot\|\bpsi_{\ell, n+1}-\bw_{k,n}\|^2
\end{align}
where $0<\tau_k\ll 1$ are forgetting factors. The evolution of the coefficient vectors $\bw_k^\star$ was divided into four stationary stages and three transient episodes.
For stationary stages, the vectors $\bw_k^\star$ were generated randomly from a standard Gaussian distribution. During stationary stages, we set $\bw_k^\star$ at each agent so that, from time instant $n = 1$ to $1000$ and $n = 4501$ to $7000$, the whole network tracked the same target, while from instant $n = 1501$ to $2500$ and $n = 3001$ to $4000$, the network started to track 2 and 3 targets, respectively. The transient episodes were designed by using linear interpolation over 500 time instants by $a+\frac{b-a}{500}(n-n_c)$, with $a$ and $b$ denoting an element of the weight of the previous and next stationary stages respectively, and $n_c$ denoting the starting instant of the current transient episode --- see~\cite{Chen2015diffusion}. Besides, we set $\varepsilon = 0.05$ and $\eta = 0.95$ for power-normalized diffusion scheme, with $\nu_{\gamma_k} = 0.01$ for static combination matrices and $\nu_{\gamma_k} = 0.04$ for adaptive combination matrices. For sign-regressor diffusion, we set $\nu_{\gamma_k}$ to $0.015$ and $0.03$ for static and adaptive combination matrices, respectively.

\begin{figure}[!t]
\begin{minipage}[b]{1.0\linewidth}
  \centering
  \centerline{\includegraphics[width=6.5cm]{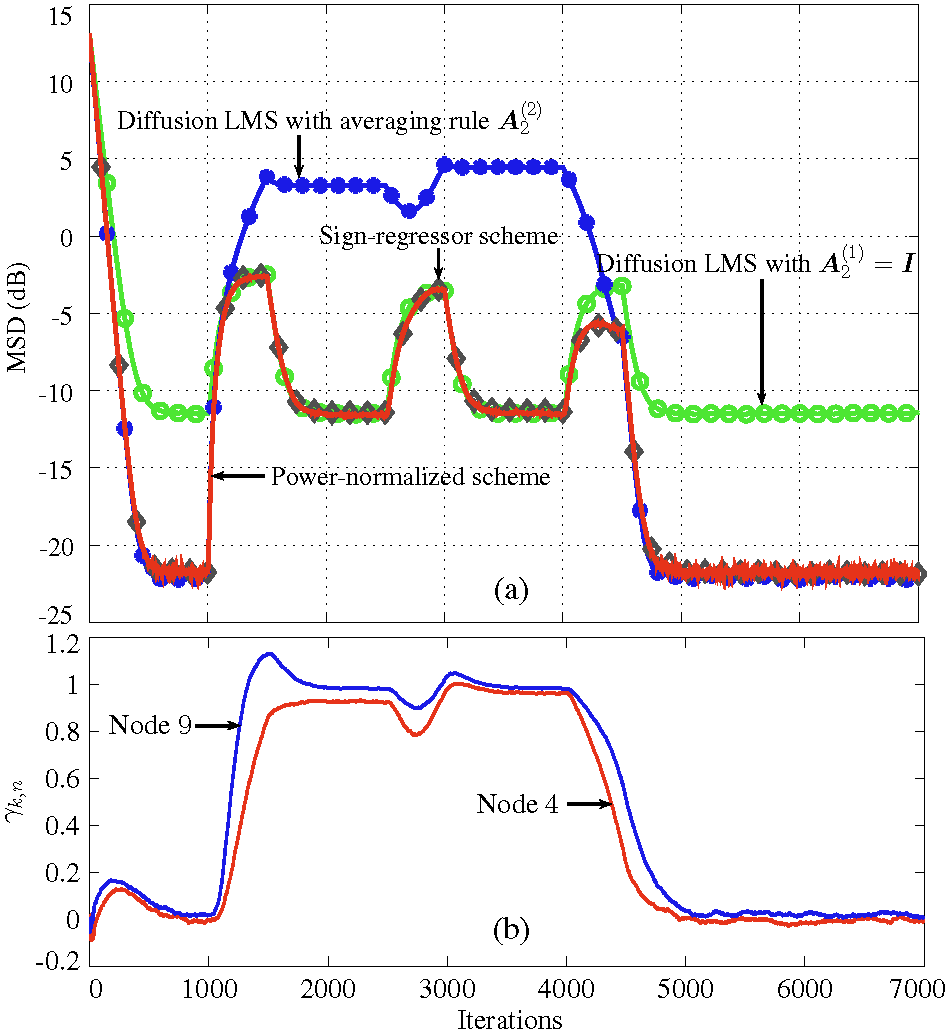}}
  \vspace{-2mm}
\end{minipage}
\caption{Simulation results with static fusion matrices. (a) Network MSD learning curves; (b) Evolution of combination coefficient $\gamma_{k,n}$ for power-normalized scheme.}
\label{fig8}
\vspace{-3mm}
\end{figure}

\begin{figure}[!t]
\begin{minipage}[b]{1.0\linewidth}
  \centering
  \centerline{\includegraphics[width=6.5cm]{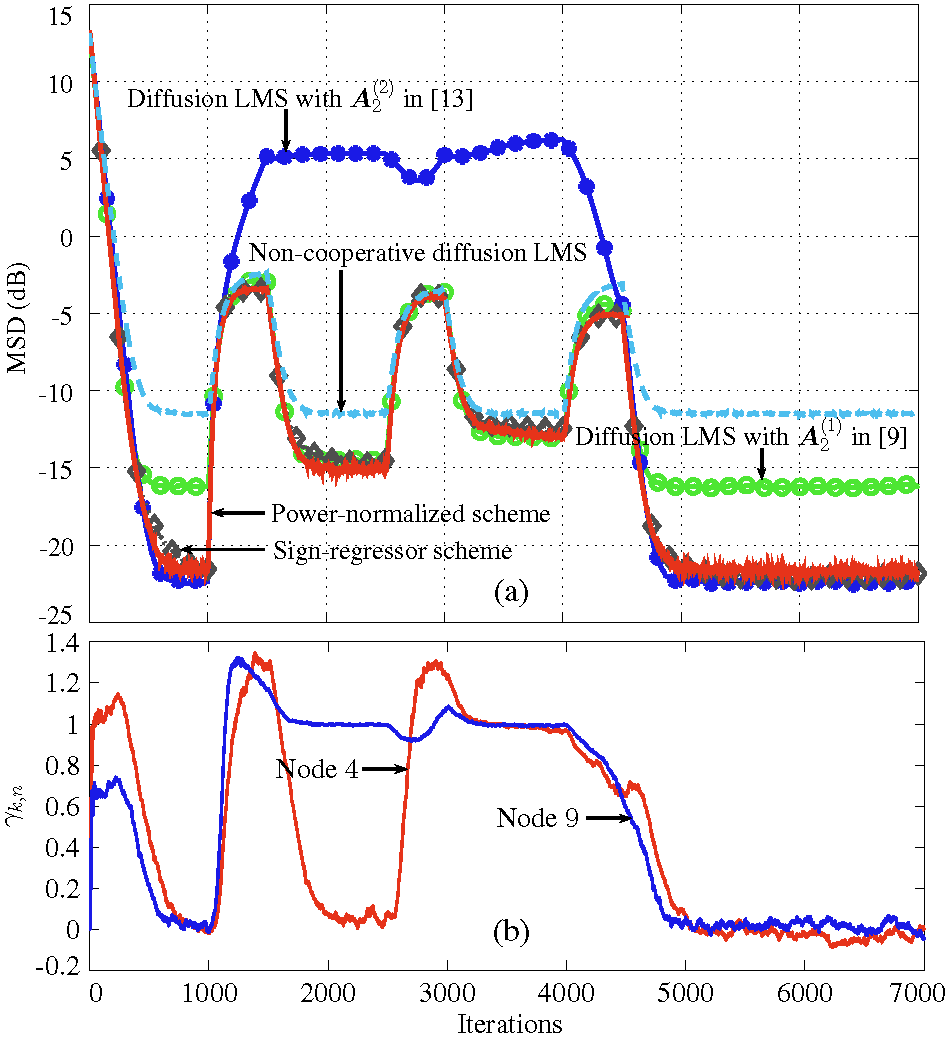}}
  \vspace{-2mm}
\end{minipage}
\caption{Simulation results with adaptive fusion matrices.}
\label{fig9}
\vspace{-5mm}
\end{figure}

The results are plotted in Figs.~\ref{fig8} and \ref{fig9} for static and adaptive fusion matrices, respectively. In Fig.~\ref{fig8}(a), as expected, the power-normalized and sign-regressor diffusion schemes led to a MSD learning curve approaching the best of each component strategies at the different stages. This coincides with the theoretical results that these schemes are universal at steady state. The evolutions of affine combination coefficients in Fig.~\ref{fig8}(b) ensure the effectiveness of proposed schemes.

The results for adaptive fusion matrices plotted in Fig.~\ref{fig9} lead to similar conclusions as Fig.~\ref{fig8}. Interestingly, from time instant $n = 1901$ to $2500$, the combined result of the power-normalized diffusion scheme outperformed each individual one, driven by combination coefficients $\gamma_{k,n}$ such as at node 4 in Fig.~\ref{fig9}(b). This confirms the theoretical result in \eqref{eq:EMSEnetconclusion} that shows that the combined strategy can outperform each component strategy in certain situations. The power-normalized diffusion performs better than the sign-regressor diffusion due to its faster convergence rate. All of the results in Fig.~\ref{fig8} and Fig.~\ref{fig9} illustrate the effectiveness of the proposed schemes.

\subsubsection{Affine combination of other strategies}\label{subsubsec:ExtensionOfCombination}

Consider the diffusion strategies for clustered multi-task networks proposed in \cite{Chen2014multitask} and \cite{Nassif2015}.
The former uses squared $\ell_2$-norm co-regularizer to promote cooperation within clusters, while the latter uses an $\ell_1$-norm co-regularizer. The simulation setting is similar to that of Section \ref{subsec:affineValidation}, except that the $10$ nodes were grouped into $3$ clusters,
to track three groups of different but related targets.
For stationary stages, the coefficient vectors $\bw^\star_{{\cal C}_i}$ were generated as $\bw^\star_{{\cal C}_i} = \bw_o + \delta_{{\cal C}_i}\bw_{{\cal C}_i}$, with $\bw_{{\cal C}_i}$ drawn from standard Gaussian distribution. When $\delta_{{\cal C}_i}$ for $i=1, 2, 3$ are the same or similar, \cite{Nassif2015} with $\ell_1$ norm co-regularizer works better, otherwise the method in \cite{Chen2014multitask} is better.
The regularization strength of the two co-regularizers were both set to $0.1$, and a uniform $\bA_2^{(i)}$ was used such that $a_{2,\ell k}^{(i)}={|{\cal N}_k\cap{\cal C}(k)|}^{-1}$. For the power-normalized diffusion scheme, we set $\nu_{\gamma_k}$ to $0.01$, $\varepsilon$ to 0.05, and $\eta$ to 0.95. For sign-regressor diffusion, we set $\nu_{\gamma_k}$ to $0.1$.

\begin{figure}[!t]
\begin{minipage}[b]{1.0\linewidth}
  \centering
  \centerline{\includegraphics[width=6.5cm]{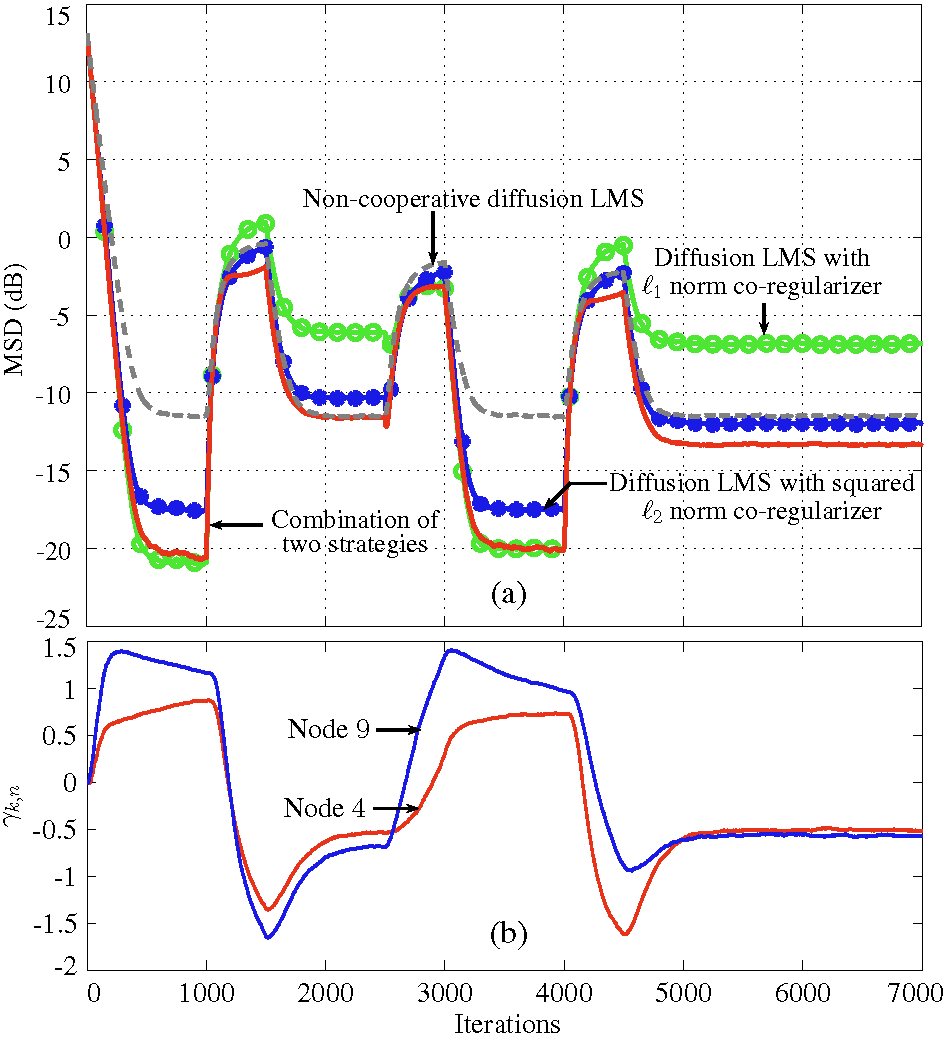}}
\end{minipage}
\caption{Simulation results of the power-normalized scheme with the two diffusion strategies in \cite{Chen2014multitask} and \cite{Nassif2015}. (a) Network MSD learning curves; (b) Evolution of the affine combination coefficients $\gamma_{k,n}$.}
\label{fig12}
\end{figure}

\begin{figure}[!t]
\begin{minipage}[b]{1.0\linewidth}
  \centering
  \centerline{\includegraphics[width=6.5cm]{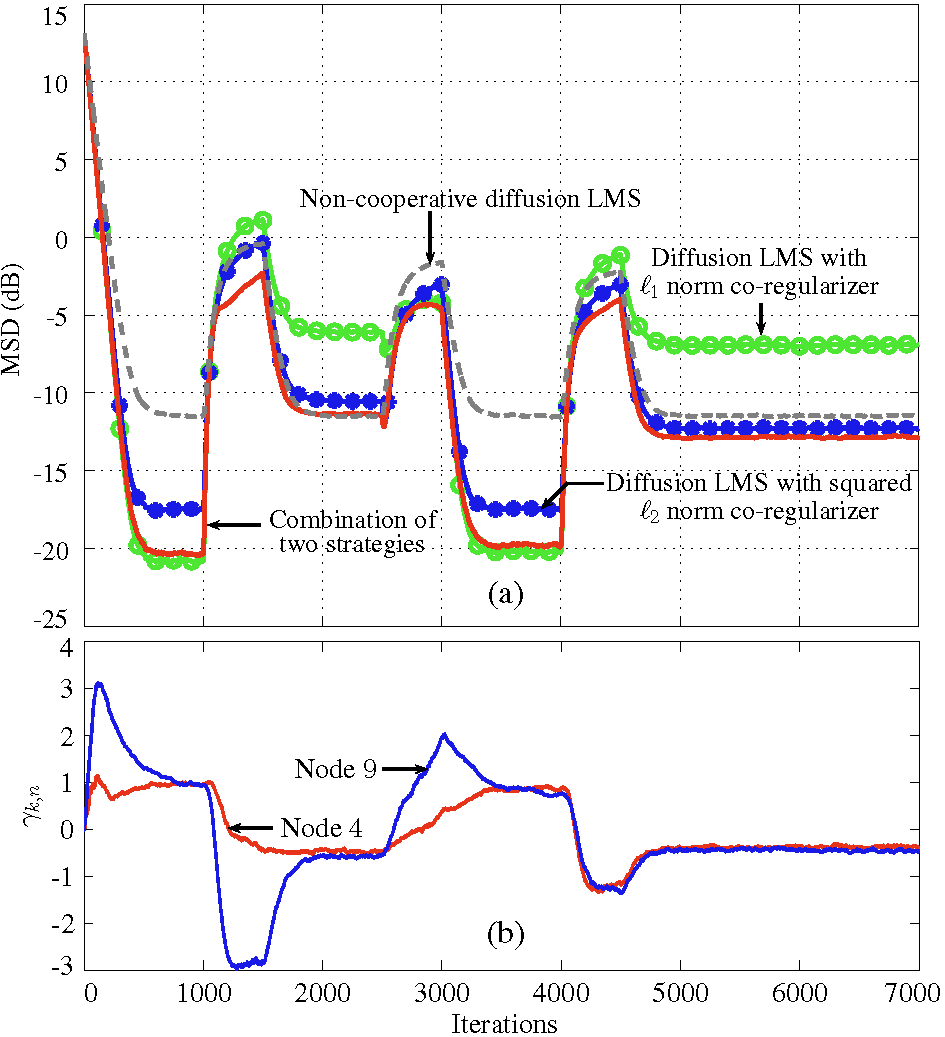}}
\end{minipage}
\caption{Simulation results of the sign-regressor scheme with the two diffusion strategies in \cite{Chen2014multitask} and \cite{Nassif2015}.}
\label{fig13}
\end{figure}

The results are plotted in Fig.~\ref{fig12} and Fig.~\ref{fig13}. Both power-normalized and sign-regressor diffusion not only led to the best of each component strategies at different stages, they also outperformed each component strategies at some instants.

\subsubsection{Influence of parameters}\label{subsubsec:Parameters}
Since there are several parameters in the power-normalized and sign-regressor schemes, such as step-size $\nu_{\gamma_{k}}$, temporal smoothing factor $\eta$ and parameter $\varepsilon$, we examine their influence on the performance. Based on various experiments, we find that the performance of the power-normalized scheme is not sensitive to $\eta$ and to small-valued $\varepsilon$. We therefore suggest setting $\eta$ to a typical value of $0.95$ and $\varepsilon$ to $0.05$. We also examine the influence of the step-size $\nu_{\gamma_{k}}$. The simulation settings are identical to those used in the first experiment, and we use static combination matrices.

\begin{figure}[!t]
\begin{minipage}[b]{1.0\linewidth}
  \centering
  \centerline{\includegraphics[width=6.5cm]{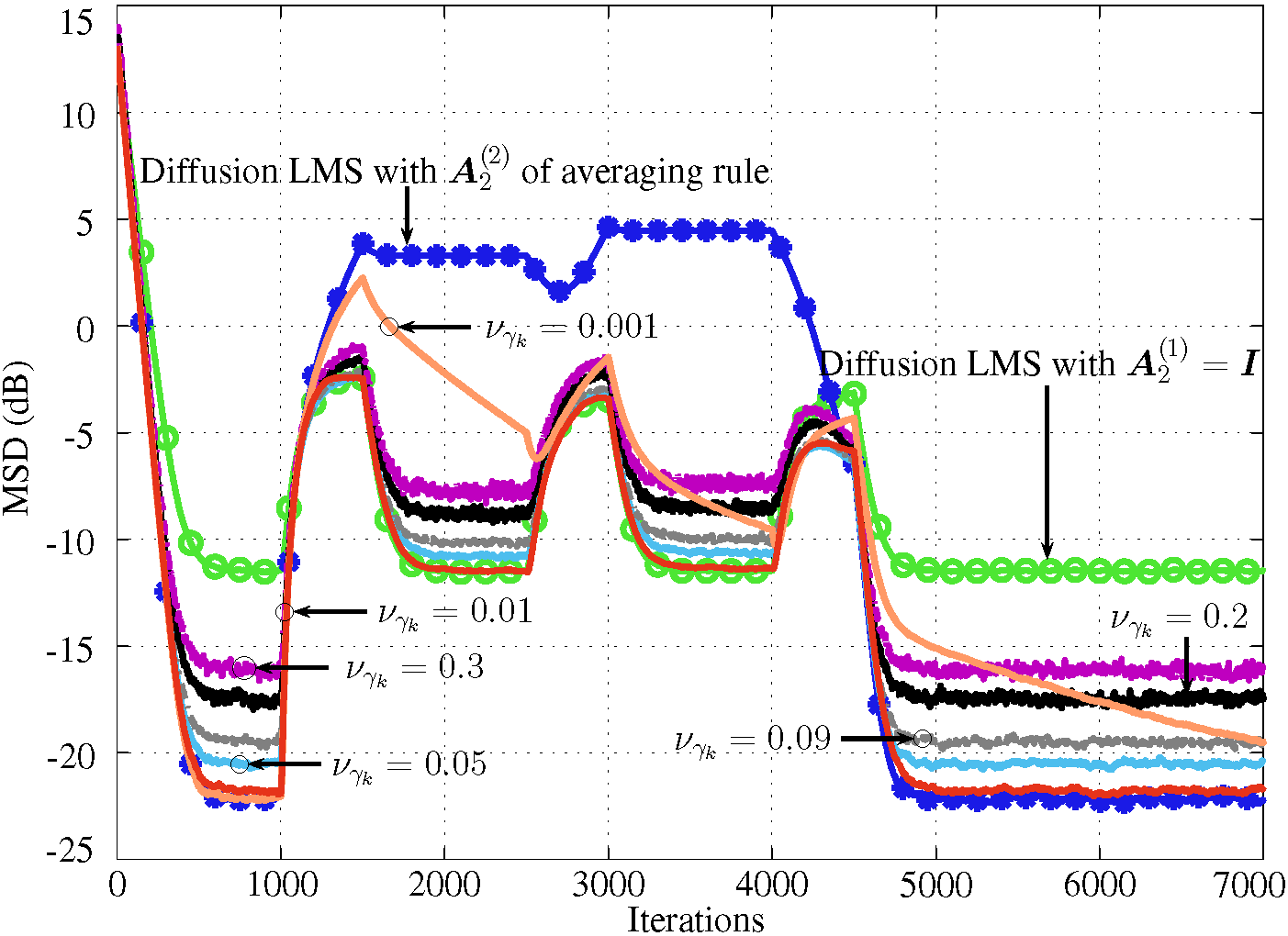}}
\end{minipage}
\caption{Simulation results of the power-normalized scheme with different step-sizes $\nu_{\gamma_{k}}$.}
\label{figR1}
\end{figure}

\begin{figure}[!t]
\begin{minipage}[b]{1.0\linewidth}
  \centering
  \centerline{\includegraphics[width=6.5cm]{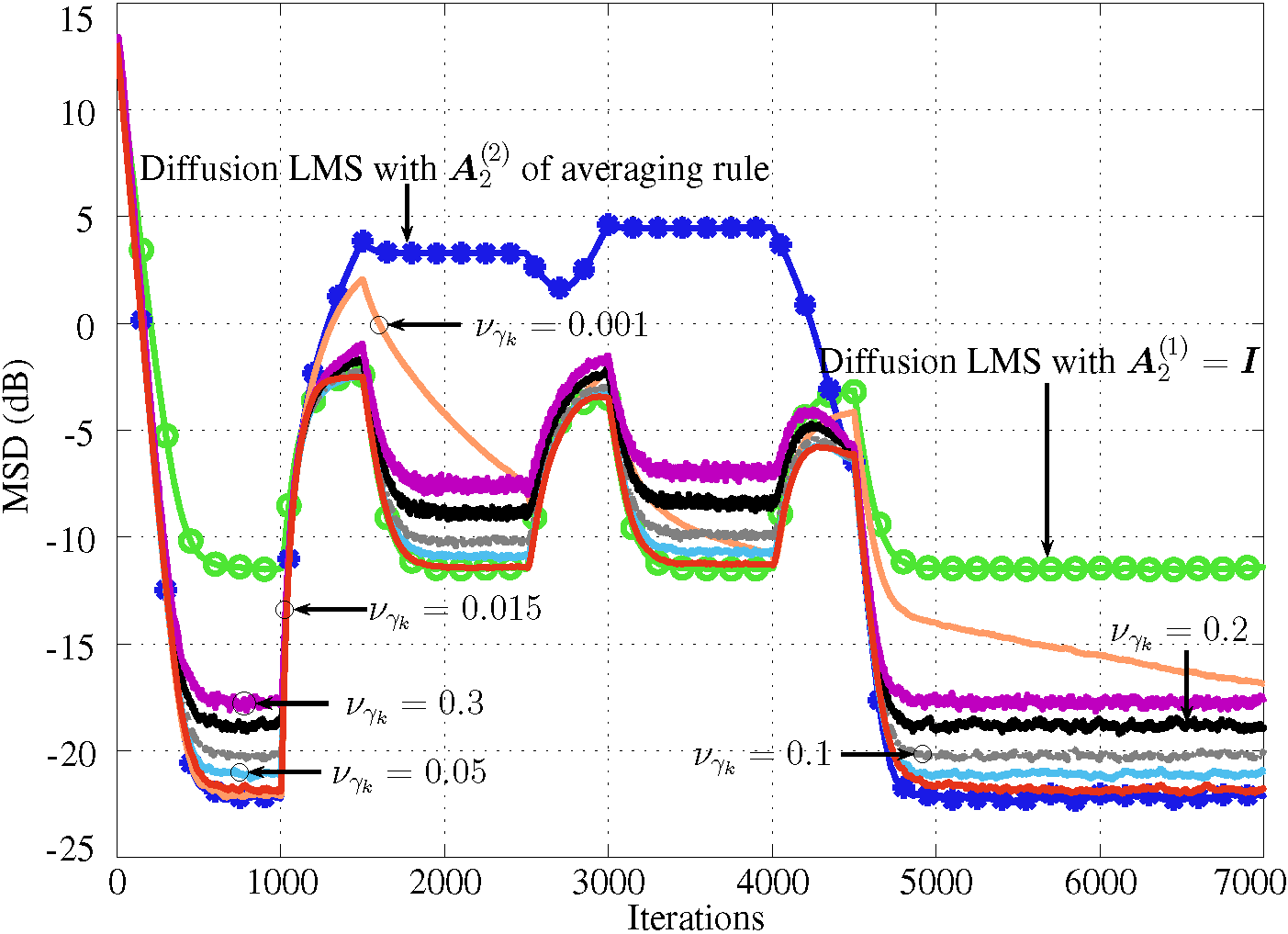}}
\end{minipage}
\caption{Simulation results of the sign-regressor scheme with different step-sizes $\nu_{\gamma_{k}}$.}
\label{figR2}
\end{figure}

The results are plotted in Figs.~\ref{figR1} and~\ref{figR2}. For both power-normalized and sign-regressor diffusion,   small $\nu_{\gamma_{k}}$ lead to weak ability in tracking the best component and slow convergence toward the best component at steady state, such as $\nu_{\gamma_k}=0.001$, while a large $\nu_{\gamma_{k}}$ results in biases from the best component at steady state, though a large step-size ensures good tracking for the best component. The step-size parameter $\nu_{\gamma_k}$ needs to be fine-tuned to ensure good tracking ability and a lower bias from the best component at steady state. In this simulation setting, the values of $\nu_{\gamma_{k}}$ are set to $0.01$ and $0.015$ for power-normalized and sign-regressor schemes, respectively.

\vspace{-4mm}

\subsection{Theoretical models}\label{subsec:Theoreticalvalidation}

To illustrate the theoretical results as well as challenge the assumptions and approximations adopted in the theoretical analysis, we considered three networks with different connectivity parameters as described in Table~\ref{NetParameters}.  \texttt{Net1} consisted of $10$ nodes with the network topology given in Fig.~2(a). \texttt{Net2} was a more complicated network consisting of $20$ nodes. \texttt{Net3} was generated by dividing $20$ nodes into seven fully connected clusters, with  $3$ nodes in each of the first six clusters and $2$ nodes in the last cluster. These seven clusters were connected in chain, with a single edge connecting adjacent clusters: agent $3$ (in cluster $1$) was connected with agent $4$ (in cluster $2$), and agent $6$ (in cluster $2$) was  connected with agent $7$ (in cluster $3$), and so on until agent $18$ (in cluster $6$) was connected to agent $19$ (in cluster $7$).

\begin{table}[!htbp]
\centering
\caption{Network statistics for theoretical models validation. $\bm{L}$ is the Laplacian matrix associated with the graph (network), $\lambda_2(\bm{L})$ is the algebraic connectivity \cite{Fiedler73} of graph, \texttt{size} is the number of nodes, \texttt{density} is the number of non-zero entries of the adjacency matrix of graph, and \texttt{diameter} is the maximum distance between any two nodes~\cite{Simoes2019FADE}.}
\label{NetParameters}
\vspace{-1mm}
\begin{tabular}{c c c c c}\hline
\texttt{Network} & \texttt{Size} & \texttt{Density} & $\lambda_2(\bm{L})$ & \texttt{Diameter} \\\hline
\texttt{Net1}& 10 & 44\% & 0.7962 & 3 \\
\texttt{Net2}& 20 & 38\%& 0.9549 & 3\\
\texttt{Net3}& 20 & 17.25\%& 0.0439 & 13\\\hline
\end{tabular}
\end{table}

The unknown coefficient vectors to be estimated were of length $L = 2$. We first considered regressors drawn from a zero-mean Gaussian distribution with covariance matrix $\bR_{x,k} = \sigma_{x,k}^2\bI_L$. Next we considered colored regressors $\bx_{k,n} = [x_{k,n}\,\,x_{k,n-1}]^\top$
generated from a first-order AR model:
$x_{k,n}=0.5x_{k,n-1}+\sqrt{0.75\sigma^2_{x,k}}w_{k,n}.$
The input signal $w_{k,n}$ was i.i.d and drawn from a zero-mean Gaussian distribution, with variance $\sigma^2_w=1$, so that:
\begin{equation}
\bR_{x,k}=\sigma^2_{x,k}\left(\begin{array}{cc}
1&0.5\\
0.5&1
\end{array}\right).
\end{equation}

\begin{figure*}[!t]
\begin{minipage}[b]{1.0\linewidth}
  \centering
  \centerline{\includegraphics[width=18cm]{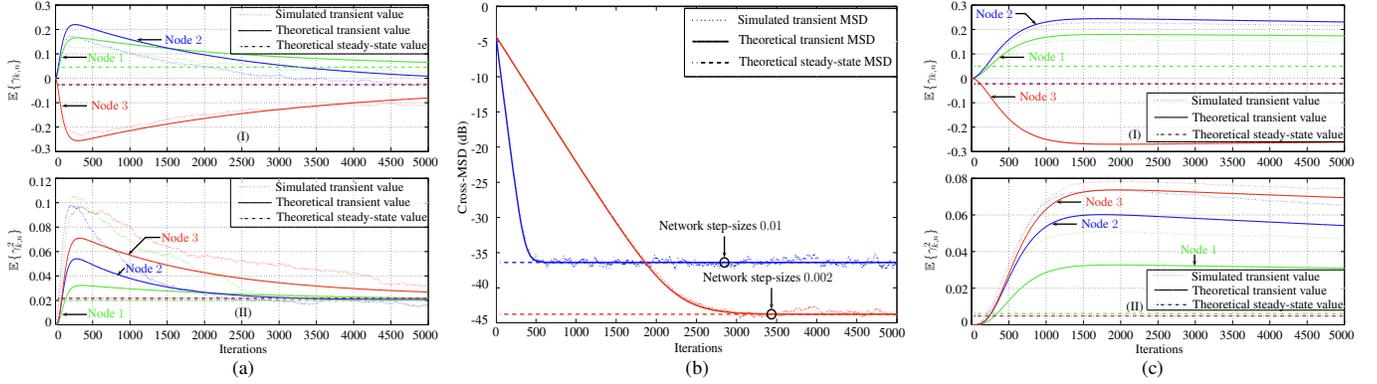}}
  \vspace{-2mm}
\end{minipage}
\caption{Illustration of simulation results (model vs. Monte Carlo) for the power-normalized scheme in \texttt{Net1} and \texttt{SNR1}. Transient and steady-state values of $\E\{\gamma_{k,n}\}$ derived in \eqref{eq:meanIterationAFF} and \eqref{eq:affStationarySolution} (\emph{top}), as well as these of $\E\{\gamma_{k,n}^2\}$ derived in \eqref{eq:AFFPNLMSExpectation} and \eqref{eq:gammasquaresteadyAff} (\emph{bottom}) for network step-size $0.01$ (a) and $0.002$ (c); (b) Transient and steady-state cross-MSDs derived in \eqref{eq:MSDIterationCross} and \eqref{eq:crossterm-steady}.}
\label{fig22}
\end{figure*}

\begin{figure}[!t]
\begin{minipage}[b]{1.0\linewidth}
  \centering
  \centerline{\includegraphics[width=6.5cm]{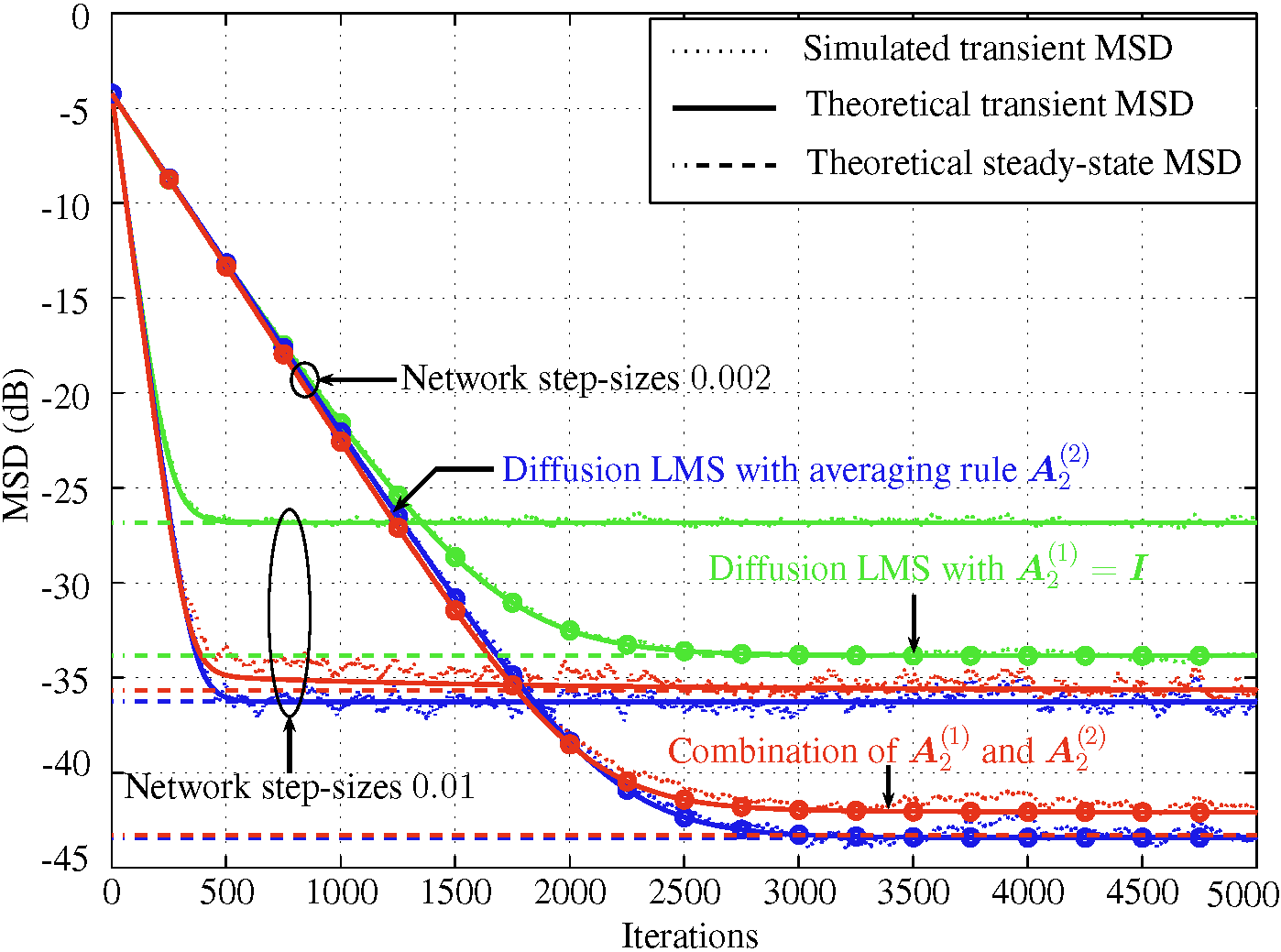}}
\end{minipage}
\caption{Network performance with the power-normalized scheme (model vs. Monte Carlo) with two different network step-sizes in \texttt{Net1} and \texttt{SNR1}. For each step-size, the results plotted with a same color stand for same diffusion strategy. For each component diffusion strategy, theoretical transient MSD and theoretical steady-state MSD are derived in \eqref{eq:MSDIteration} and \eqref{eq:steadyMSD.iterationi}. The combination results are derived in \eqref{eq:meansquareanalysis} and \eqref{eq:netSteadyMSD}.}
\label{fig23}
\end{figure}

\begin{figure*}[!t]
\begin{minipage}[b]{1.0\linewidth}
  \centering
  \centerline{\includegraphics[width=18cm]{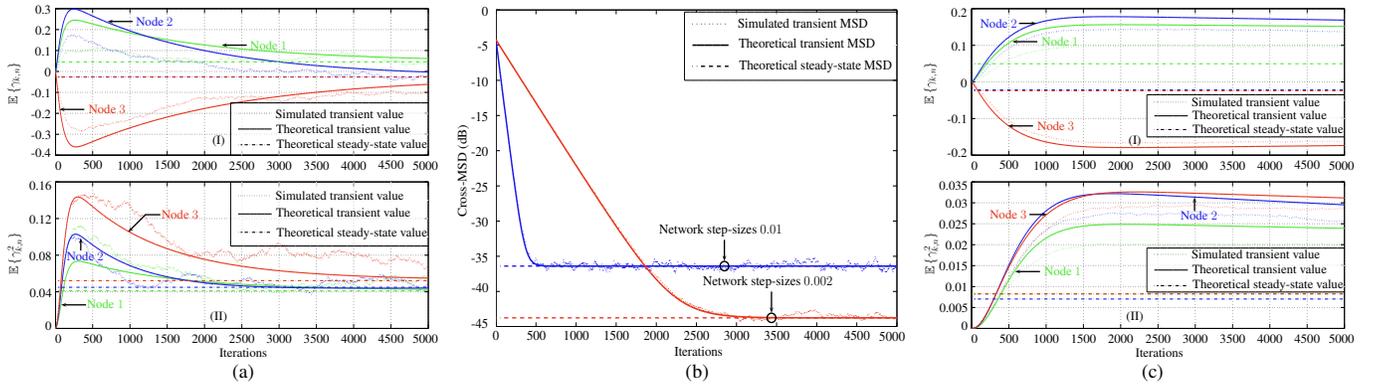}}
\end{minipage}
\caption{Illustration of simulation results (model vs. Monte Carlo) of the sign-regressor scheme in \texttt{Net1} and \texttt{SNR1}. Transient and steady-state values of $\E\{\gamma_{k,n}\}$ derived in \eqref{eq:IterTheoreticalAffSRresult} and \eqref{eq:affStationarySolution} (\emph{top}), as well as $\E\{\gamma_{k,n}^2\}$ derived in \eqref{eq:adaptAffSRmeansquareexpect} and \eqref{eq:gammasquaresteadyAff} (\emph{bottom}) for network step-size $0.01$ (a) and $0.002$ (c); (b) Transient and steady-state cross-MSDs derived in \eqref{eq:MSDIterationCross} and \eqref{eq:crossterm-steady}.}
\label{fig24}
\end{figure*}

The noise signals were generated from Gaussian distributions $\cp{N}(0,\sigma_{z,k}^2)$. Variances $\sigma_{x,k}^2$ and $\sigma_{z,k}^2$ at each agent were generated randomly. By varying $\sigma_{z,k}^2$, we changed the signal-to-noise ratio (SNR) \cite{Das2013Distributed} to three levels as described in Table~\ref{SNRParameters}. Both combination schemes were run with network step-sizes being set to $\{0.01,\,0.002\}$, and with $\nu_{\gamma_k}$ correspondingly being set to $\{0.01,\,0.002\}$ and $\{0.015,\,0.001\}$ for power-normalized diffusion and sign-regressor diffusion, respectively. Besides, we set $\varepsilon$ to $0.05$ and $\eta$ to $0.95$ for power-normalized diffusion. We first validated the theoretical results related to the mean and mean-square behaviors of $\gamma_{k,n}$, the theoretical MSD of each component strategy, as well as cross-MSD of the whole network defined by ${\rm MSD}_{\rm cross} \triangleq \frac{1}{N}\E\bigl\{\bv_{n}^{(1)\top}\bv_{n}^{(2)}\bigr\}$. Then we evaluated the theoretical MSD behavior of the combined strategy.

\begin{table}[!htbp]
\centering
\caption{Three SNR levels in decibel (dB) for theoretical models validation. Since SNRs vary from node to node, we enumerate the {Maximum}, {Minimum} and {Mean} values.}
\label{SNRParameters}
\vspace{-1mm}
\begin{tabular}{c c c c}\hline
{SNR Level} & {Maximum} & {Minimum} & {Mean} \\\hline
\texttt{SNR1}& 3.5724 & 1.6673 & 2.7946 \\
\texttt{SNR2}& -9.4379 & -11.343& -10.2157 \\
\texttt{SNR3}& -18.9803 & -20.8855& -19.7581\\\hline
\end{tabular}
\vspace{-2mm}
\end{table}

\begin{figure}[!t]
\begin{minipage}[b]{1.0\linewidth}
  \centering
  \centerline{\includegraphics[width=6.5cm]{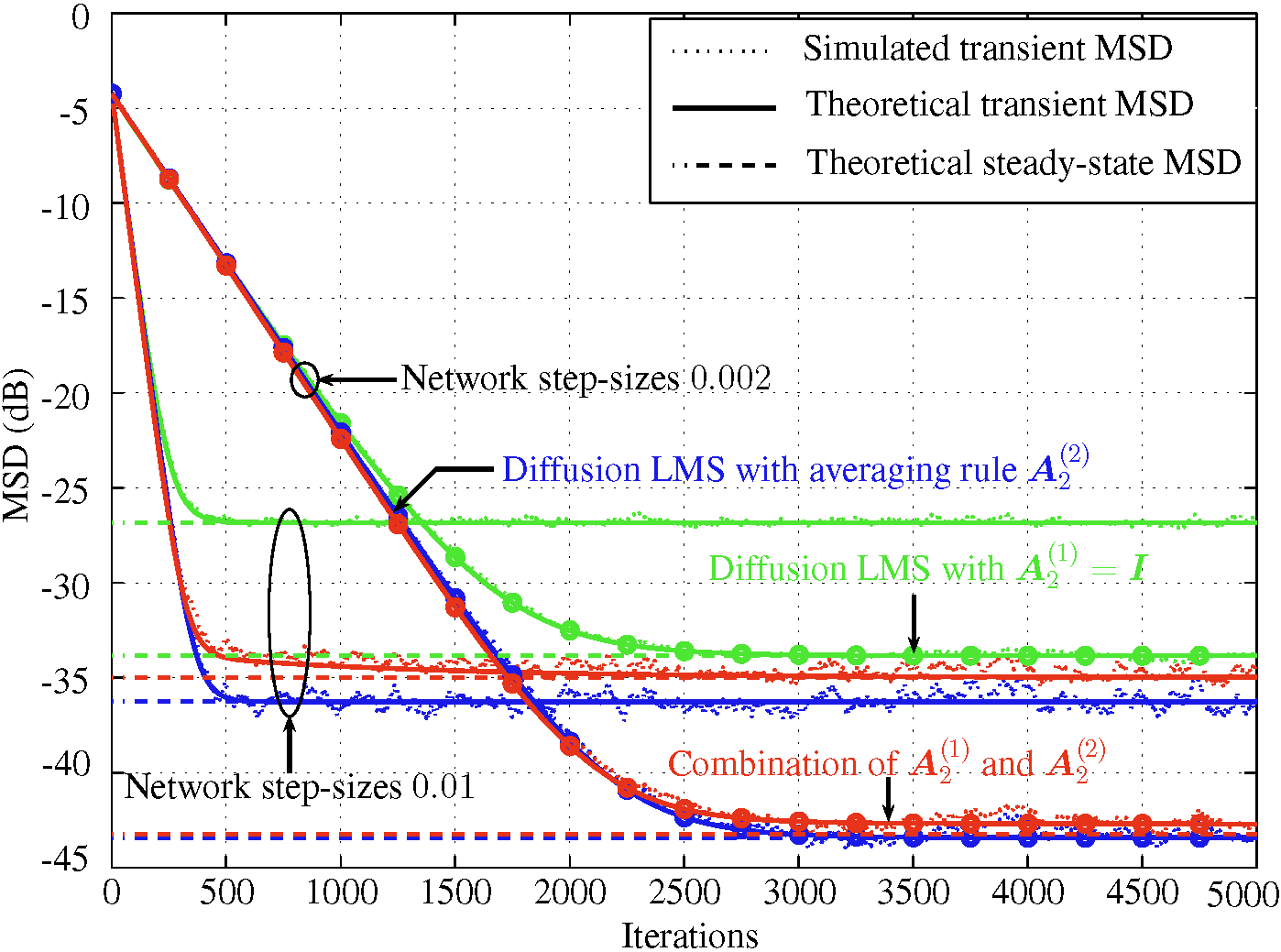}}
  \vspace{-2.5mm}
\end{minipage}
\caption{Network performance with the sign-regressor scheme in \texttt{Net1} and \texttt{SNR1}.}
\label{fig25}
\vspace{-3mm}
\end{figure}

The results for white Gaussian inputs with \texttt{Net1} and \texttt{SNR1} are plotted in Fig.~\ref{fig22}--Fig.~\ref{fig25}. In both Fig.~\ref{fig22} and Fig.~\ref{fig24}, we observe that the simulated transient values and theoretical transient values accurately matched, especially for small step-sizes, since the approximations adopted in theoretical analyses are more reasonable for small step-sizes. Further, we observe that a larger $\nu_{\gamma_k}$ results in a faster convergence rate, which coincidences with equations \eqref{eq:ConverRadiusPN} and \eqref{eq:AffSRstepsize} in theoretical analyses. In Fig.~\ref{fig23} and Fig.~\ref{fig25}, besides the accurate matching of the simulated and theoretical MSD learning curves, the almost superposition of theoretical steady-state MSDs for combined strategy and that of the best component validates the conclusion again that the combination schemes are universal at steady state. Further, the gaps of these two theoretical values becomes smaller as the step-sizes decrease, since the analyses of the universality are based on the approximations that are more valid for small step-sizes.

\begin{figure}[!t]
\begin{minipage}[b]{1.0\linewidth}
  \centering
  \centerline{\includegraphics[width=6.5cm]{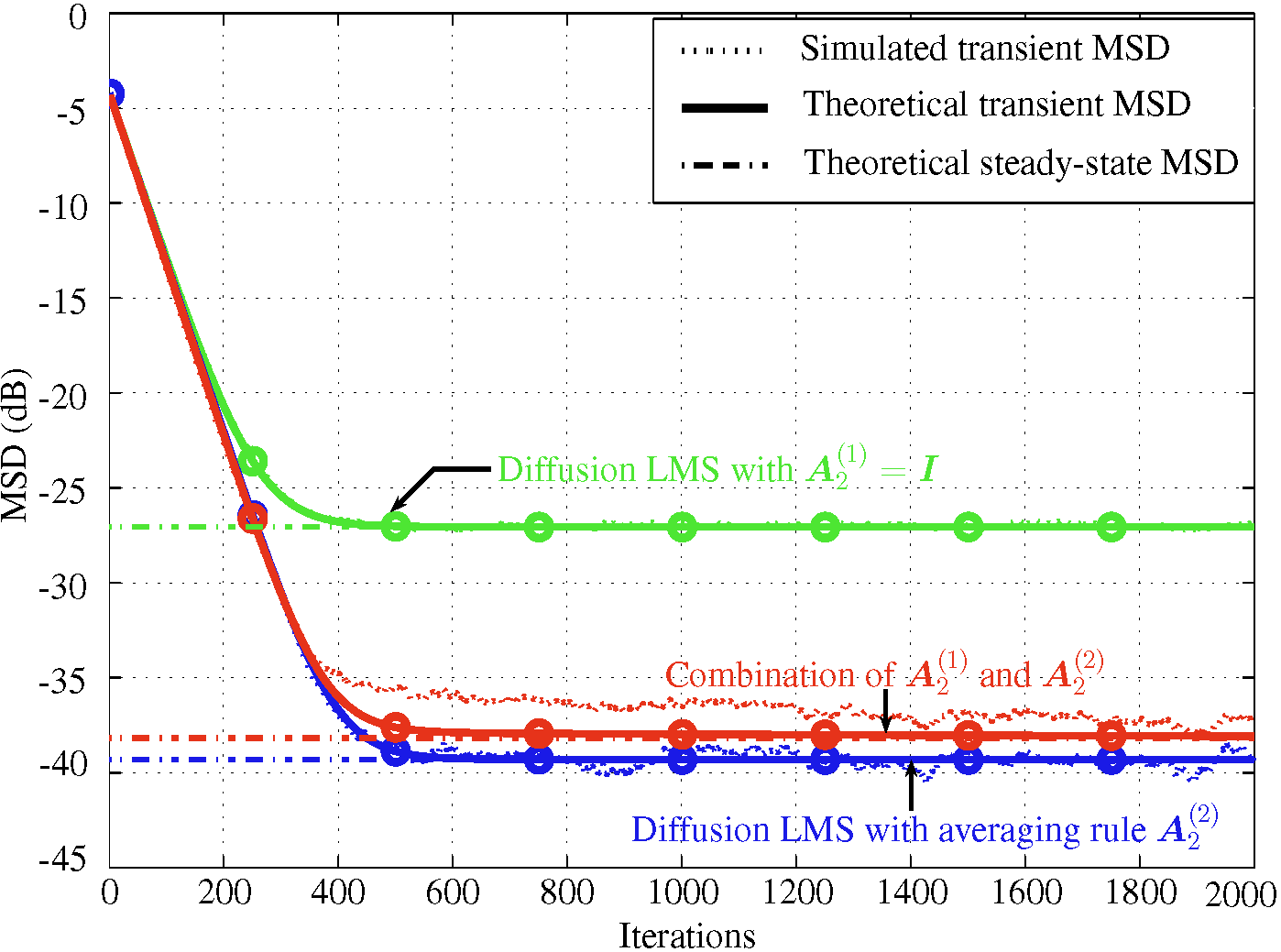}}
  \vspace{-2.5mm}
\end{minipage}
\caption{Network performance with the power-normalized scheme in \texttt{Net2} and \texttt{SNR1}, and the network step-sizes are $0.01$.}
\label{figNet2}
\end{figure}

\begin{figure}[!t]
\begin{minipage}[b]{1.0\linewidth}
  \centering
  \centerline{\includegraphics[width=6.5cm]{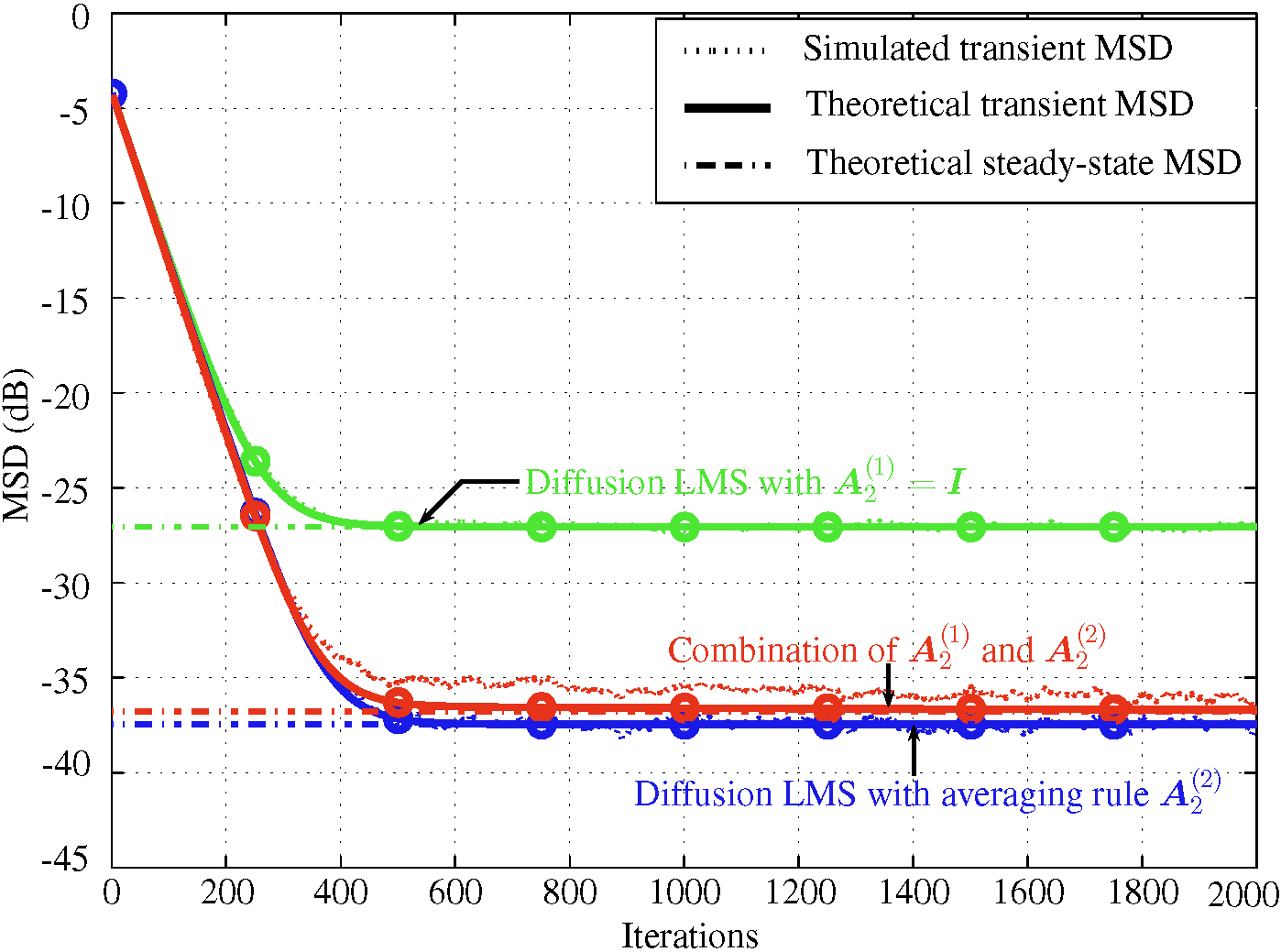}}
\end{minipage}
\caption{Network performance with the power-normalized scheme in \texttt{Net3} and \texttt{SNR1}, and the network step-sizes are $0.01$}.
\label{figNet3}
\end{figure}

\begin{figure}[!t]
\begin{minipage}[b]{1.0\linewidth}
  \centering
  \centerline{\includegraphics[width=6.5cm]{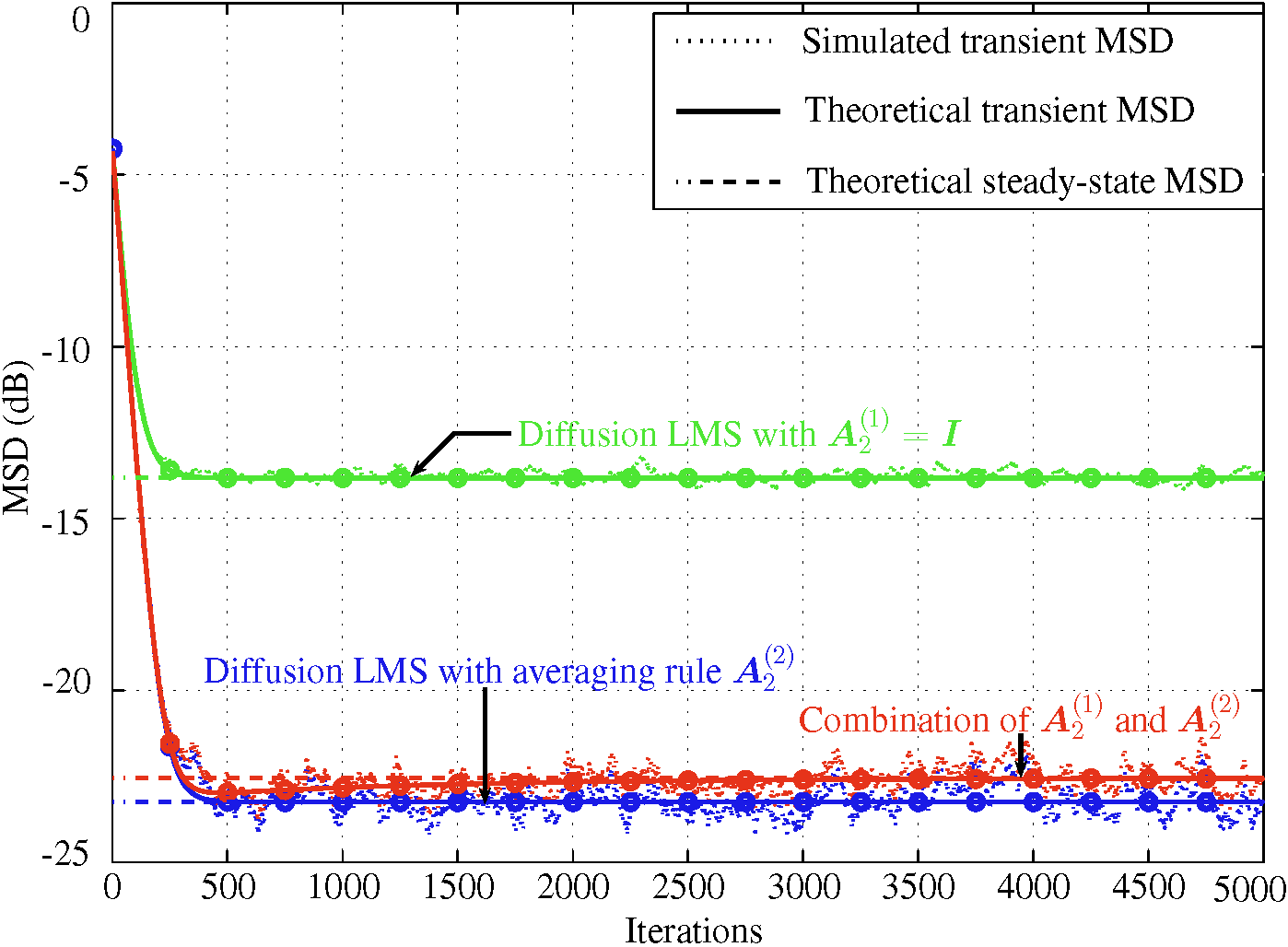}}
\end{minipage}
\caption{Network performance with the power-normalized scheme in \texttt{Net1} and \texttt{SNR2}, and the network step-sizes are $0.01$.}
\label{figSNR-10}
\end{figure}

\begin{figure}[!t]
\begin{minipage}[b]{1.0\linewidth}
  \centering
  \centerline{\includegraphics[width=6.5cm]{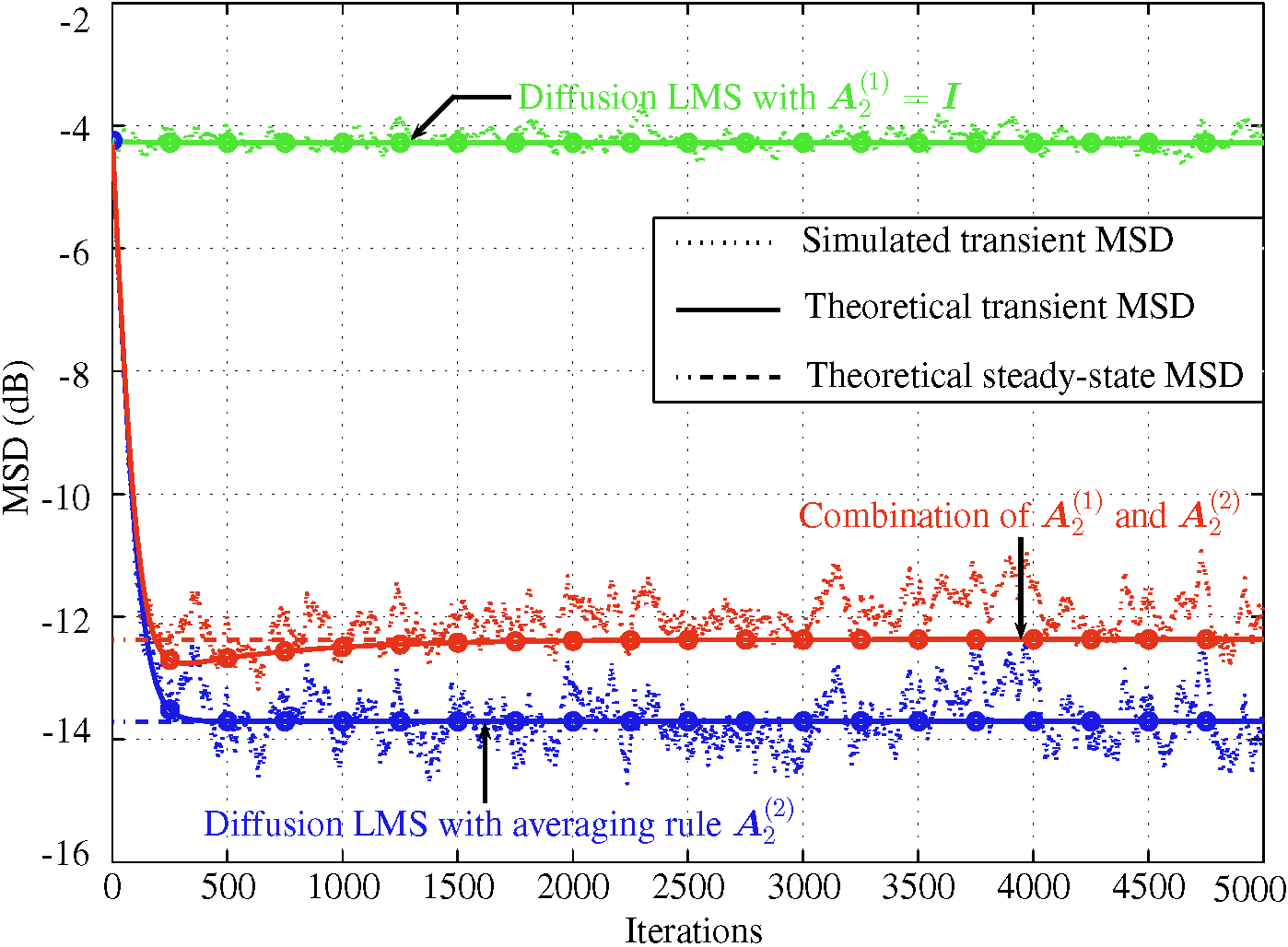}}
\end{minipage}
\caption{Network performance with the power-normalized scheme in \texttt{Net1} and \texttt{SNR3}, and the network step-sizes are $0.01$.}
\label{figSNR-20}
\end{figure}

The results of the power-normalized diffusion for white Gaussian inputs with \texttt{Net2} and \texttt{Net3} in \texttt{SNR1} are plotted in Fig.~\ref{figNet2} and Fig.~\ref{figNet3}, respectively. And the results with \texttt{Net1} in \texttt{SNR2} and \texttt{SNR3} are plotted in Fig.~\ref{figSNR-10} and Fig.~\ref{figSNR-20}, respectively. Together with Fig.~\ref{fig23}, all these validate the accuracy of theoretical analyses under different SNR conditions and network connectivity parameters.

\begin{figure}[!t]
\begin{minipage}[b]{1.0\linewidth}
  \centering
  \centerline{\includegraphics[width=6.5cm]{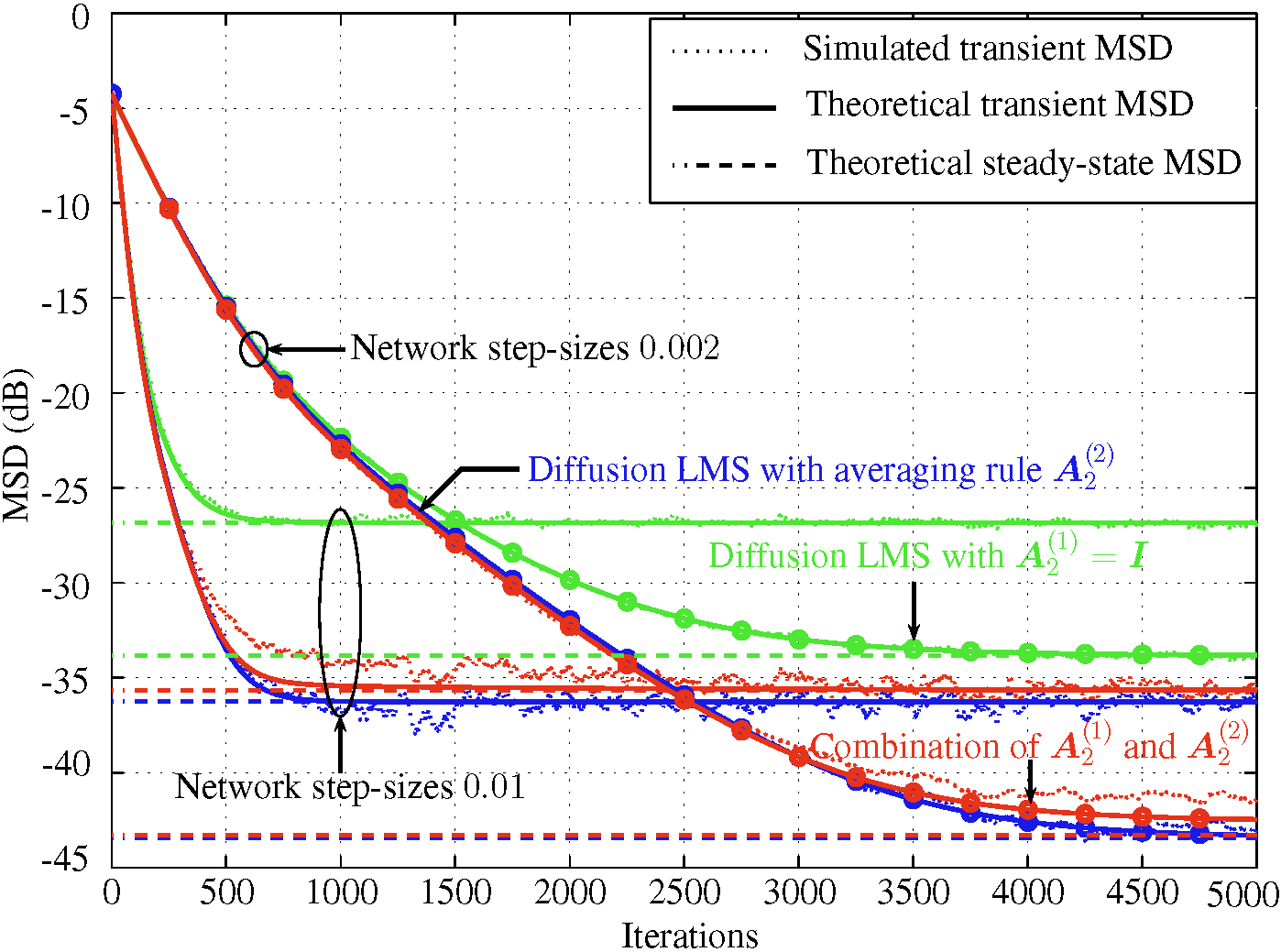}}
  \vspace{-2.5mm}
\end{minipage}
\caption{Network performance with the power-normalized scheme using correlated inputs in \texttt{Net1} and \texttt{SNR1}.}
\label{fig26}
\vspace{-4mm}
\end{figure}

The results of the power-normalized diffusion for correlated input with \texttt{Net1} and \texttt{SNR1} are plotted in Fig.~\ref{fig26}. Though assumption \textbf{A2} is violated, the superimposition of simulated and theoretical curves validates the accuracy of the theoretical analyses for sufficiently small step-sizes with moderately correlated regressors.

\section{CONCLUSION AND PERSPECTIVES}
\label{sec:conclusions}
Combining diffusion strategies enables a network to reach better performance. In this paper, we proposed two schemes for affine combination of two diffusion strategies. By using the proposed combination schemes, we obtained a combined diffusion strategy ensuring the advantages of both component strategies simultaneously, sometimes even better, in terms of EMSE. We conducted theoretical analyses in the mean and mean-square sense, and analyzed the universality of each approach. Simulation results illustrate the interesting properties of affine combination schemes, as well as the accuracy of the theoretical results. Several open problems still have to be addressed. For instance, it would be interesting to design combination schemes and conduct theoretical analysis for colored measurement noise and correlated regressors. Some works focusing on adaptive networks with colored measurement noise include \cite{Bertrand2011,Piggott2016TSP,Piggott2017TSP}. It would also be interesting to explore other combination frameworks and schemes.

\vspace{4mm}
\section*{Appendix A\\Combination Scheme for Multiple Strategies}
The general scheme follows the description from Section III except that we now have $M$ component diffusion strategies. We introduce $M$ affine combination coefficients $\gamma_{k,n}^{(1)}, \gamma_{k,n}^{(2)}, \cdots, \gamma_{k,n}^{(M)}$ for node $k$ at time instant $n$, satisfying the constraint $\sum_{i=1}^M \gamma_{k,n}^{(i)} = 1$. By combining the estimates of $M$ component strategies at each agent $k$, we obtain the overall system coefficients $\bw_{k,n}$ and estimation error $e_{k,n}$ at combination layer, with $e_{k,n} = \sum_{i=1}^M \gamma_{k,n}^{(i)} e_{k,n}^{(i)}$.
In order to keep $\gamma_{k,n}^{(i)}$ satisfying the sum-to-one constraint, we calculate $\gamma_{k,n}^{(i)}$ via the mapping:
\vspace{-2mm}
\begin{equation}
	\label{eq:mapping}
	\gamma_{k,n}^{(i)} = \frac{\alpha_{k,n}^{(i)}+\delta}{\sum_{j=1}^M \alpha_{k,n}^{(j)}+M\delta},
\vspace{-2mm}
\end{equation}
where $\alpha_{k,n}^{(i)}$ are newly introduced auxiliary parameters, and $\delta$ is a small positive constant to avoid zero-division. We shall update $\alpha_{k,n}^{(i)}$ instead of updating $\gamma_{k,n}^{(i)}$ directly. By minimizing the MSE \eqref{eq:netMSE} over the entire network,
and using stochastic gradient approximation, we obtain:
\begin{align}
{\alpha}_{k,n+1}^{(i)} 
& \approx {\alpha}_{k,n}^{(i)}+\nu_{\alpha_k}e_{k,n}\sgn\biggl\{\frac{e_{k,n} - e_{k,n}^{(i)}}{\sum_{j=1}^M \alpha_{k,n}^{(j)}+M\delta}\biggr\}.\label{eq:UpdateMS}
\end{align}

\section*{Appendix B\\Proof of Universality Analysis Result 1}
\label{sec:appendix-B}

Starting from the update equation of $\gamma_{k,n}$ in \eqref{eq:affadapta}, we know that the stationary point of $\gamma_{k,n}$ is reached for:
\begin{align}
\E\Bigl\{{\nu_{\gamma_k}\over \varepsilon + p_{k,n}}\,e_{k,n}\,\bx_{k,n}^{\top}\bigl({\bw}_{k,n}^{(1)} - {\bw}_{k,n}^{(2)}\bigl)\Bigr\}=0.\label{eq:affStationary}
\end{align}
Using \eqref{eq:eakndiffer} and \eqref{eq:equivalentekn}, we obtain:
\begin{align}
&\E\Bigl\{{\nu_{\gamma_k}\over \varepsilon + p_{k,n}}\,\gamma_{k,n}\, \bigl[\widetilde{e}_{k,n}^{(1)}\widetilde{e}_{k,n}^{(2)}
- {(\widetilde{e}_{k,n}^{(1)})}^2\bigr]+\notag\\
&{\nu_{\gamma_k}\over \varepsilon + p_{k,n}}\,{\bigl(1-\gamma_{k,n}\bigl)}\, \bigl[{(\widetilde{e}_{k,n}^{(2)})}^2 - \widetilde{e}_{k,n}^{(1)}\widetilde{e}_{k,n}^{(2)}	
\bigr]\Bigr\}=0,\label{eq:optimalconditionaff}
\end{align}
where we used the zero-mean property of $z_{k,n}$ under assumption \textbf{A1}. Though we cannot obtain a closed-form solution for $\E\{\gamma_{k,n}\}$ at each time instant, by resorting to approximations \textbf{Ap}$_1$, \textbf{Ap}$_2$ and introducing the quantity $\triangle J_{k,n}^{(i)} \triangleq J_{\text{ex},k,n}^{(i)} - J_{\text{ex},k,n}^{(1,2)}$, with $J_{\text{ex},k,n}^{(1,2)}$ denoting the cross-EMSE defined by $J_{\text{ex},k,n}^{(1,2)} \triangleq \E\{\widetilde{e}_{k,n}^{(1)}\widetilde{e}_{k,n}^{(2)}\}$, equation \eqref{eq:optimalconditionaff} at steady-state becomes:
\begin{align}
-\bar\gamma_{k,\infty}\, \triangle J_{k,\infty}^{(1)} + {\bigl(1-\bar\gamma_{k,\infty}\bigl)}\, \triangle J_{k,\infty}^{(2)}=0
\end{align}
with $\bar\gamma_{k,\infty}\triangleq\E\{\gamma_{k,\infty}\}$. This leads to a closed-form expression for the stationary point at steady-state:
\begin{align}
\bar\gamma_{k,\infty} = \frac{\triangle J_{k,\infty}^{(2)}}{\triangle J_{k,\infty}^{(1)}+\triangle J_{k,\infty}^{(2)}}.\label{eq:affStationarySolution}
\end{align}
We can now establish the universality of the power-normalized scheme \eqref{eq:affadapta} at steady-state by using \eqref{eq:affStationarySolution}. Under approximation \textbf{Ap}$_1$, and assuming further that the variance of $\gamma_{k,\infty}$ is small enough so that the approximation $\E\{\gamma_{k,\infty}^2\}\approx \bar\gamma_{k,\infty}^2$ holds, the EMSE at node $k$ after combination at steady-state can be written as:
\vspace{-2mm}
\begin{align}
	J_{\text{ex},k,\infty}
&\approx \bar\gamma_{k,\infty}^2\,J_{\text{ex},k,\infty}^{(1)} + {(1 - \bar\gamma_{k,\infty})}^2\,J_{\text{ex},k,\infty}^{(2)}\notag\\
&\quad + 2\bar\gamma_{k,\infty}{(1 - \bar\gamma_{k,\infty})}\,J_{\text{ex},k,\infty}^{(1,2)}.\label{eq:EMSEkaff}
\end{align}
By substituting $\bar\gamma_{k,\infty}$ in \eqref{eq:affStationarySolution} into \eqref{eq:EMSEkaff}, and after some algebraic manipulations, we arrive at:
\vspace{-2.5mm}
\begin{align}
	J_{\text{ex},k,\infty} = J_{\text{ex},k,\infty}^{(1,2)}+\frac{\triangle J_{k,\infty}^{(1)}\triangle J_{k,\infty}^{(2)}}{\triangle J_{k,\infty}^{(1)}+\triangle J_{k,\infty}^{(2)}}.\label{eq:EMSEkequivalentAff}
\end{align}
From Cauchy-Schwartz inequality, we know that $J_{\text{ex},k,\infty}^{(1,2)}$ cannot be simultaneously larger than $J_{\text{ex},k,\infty}^{(1)}$ and $J_{\text{ex},k,\infty}^{(2)}$. Therefore, to evaluate \eqref{eq:EMSEkequivalentAff}, we split the problem into three cases according to the relations between $J_{\text{ex},k,\infty}^{(1,2)}$ and $J_{\text{ex},k,\infty}^{(i)}$:
\begin{itemize}
\item Case 1: $J_{\text{ex},k,\infty}^{(i)} \geq J_{\text{ex},k,\infty}^{(1,2)}$ for $i = 1, 2$, which means $\triangle J_{k,\infty}^{(1)}\geq 0$ and $\triangle J_{k,\infty}^{(2)}\geq 0$ at the same time. Since $\frac{\triangle J_{k,\infty}^{(1)}\triangle J_{k,\infty}^{(2)}}{\triangle J_{k,\infty}^{(1)}+\triangle J_{k,\infty}^{(2)}}\leq\triangle J_{k,\infty}^{(1)}$ and
$\frac{\triangle J_{k,\infty}^{(1)}\triangle J_{k,\infty}^{(2)}}{\triangle J_{k,\infty}^{(1)}+\triangle J_{k,\infty}^{(2)}}\leq\triangle J_{k,\infty}^{(2)}$, we conclude from \eqref{eq:EMSEkequivalentAff} that:
\begin{align}
	J_{\text{ex},k,\infty} \leq J_{\text{ex},k,\infty}^{(1)}\quad \text{and}\quad J_{\text{ex},k,\infty}\leq J_{\text{ex},k,\infty}^{(2)}.\label{eq:EMSEkconclusionAffsingle}
\end{align}

\item Case 2: $J_{\text{ex},k,\infty}^{(1)} < J_{\text{ex},k,\infty}^{(1,2)} < J_{\text{ex},k,\infty}^{(2)}$, which corresponds to $\triangle J_{k,\infty}^{(1)}< 0$ and $\triangle J_{k,\infty}^{(2)}> 0$. Then \eqref{eq:EMSEkequivalentAff} leads to:
\vspace{-2mm}
\begin{align}
	J_{\text{ex},k,\infty} = J_{\text{ex},k,\infty}^{(1)}-\frac{{\bigl(\triangle J_{k,\infty}^{(1)}\bigr)}^{2}}{\triangle J_{k,\infty}^{(1)}+\triangle J_{k,\infty}^{(2)}}.\label{eq:AffCase1}
\vspace{-2mm}
\end{align}
Since ${\triangle J_{k,\infty}^{(1)}+\triangle J_{k,\infty}^{(2)}}> 0$,
we have $J_{\text{ex},k,\infty} < J_{\text{ex},k,\infty}^{(1)}$. As a result, $J_{\text{ex},k,\infty} < \min\,\bigl\{J_{\text{ex},k,\infty}^{(1)}, J_{\text{ex},k,\infty}^{(2)}\bigr\}$.

\item Case 3: $J_{\text{ex},k,\infty}^{(1)} > J_{\text{ex},k,\infty}^{(1,2)} > J_{\text{ex},k,\infty}^{(2)}$. Due to the symmetry of \eqref{eq:EMSEkequivalentAff} for $\triangle J_{k,\infty}^{(1)}$ and $\triangle J_{k,\infty}^{(2)}$, we obtain the same conclusion as Case 2.
\end{itemize}

Therefore, at steady-state when the combination coefficient \eqref{eq:affStationarySolution} is applied, the resulting EMSE after combination cannot be worse than that of the best component, that is,
\begin{align}
	J_{\text{ex},k,\infty} \leq \min\,\bigl\{J_{\text{ex},k,\infty}^{(1)}, J_{\text{ex},k,\infty}^{(2)}\bigr\}.\label{eq:EMSEkconclusionAff}
\end{align}
We also conclude that equation \eqref{eq:EMSEnetconclusion} holds. It means that the EMSE of the diffusion network after combination cannot be worse than that of the best component strategies, leading to the universality property of the power-normalized scheme at steady state.
\vspace{-2mm}

\section*{Appendix C\\Proof of Stability Analysis Result 1}
\label{sec:appendix-C}

$\E\bigl\{\bv_{n+1}\bigr\}$ converges as $n\to\infty$ if, and only if, both terms on the right-hand side (RHS) of \eqref{eq:bvnfinal} converge to finite values. Using \eqref{eq:compactbvstarexpect2} and iterating the first term from $\E\bigl\{\bv^{(1)}_0\bigr\}$, we obtain:
\vspace{-2mm}
\begin{align}
\E\bigl\{\bGamma_{n+1}\bigr\}\,\E\bigl\{\bv_{n+1}^{(1)}\bigr\} & = \E\bigl\{\bGamma_{n+1}\bigr\}{\bigl(\overline\bB^{(1)}\bigr)}^{n+1}\E\bigl\{\bv^{(1)}_0\bigr\}\notag\\
&\quad - \sum_{j=0}^n\E\bigl\{\bGamma_{n+1}\bigr\}{\bigl(\overline\bB^{(1)}\bigr)}^{j}\overline\br^{(1)}.\label{eq:meanProve}
\end{align}
To prove convergence of \eqref{eq:meanProve}, it is sufficient to prove that the sequence $s_{1,m}\triangleq\Bigl[\E\bigl\{\bGamma_{n+1}\bigr\}{\bigl(\overline\bB^{(1)}\bigr)}^{n+1}\E\bigl\{\bv^{(1)}_0\bigr\}\Bigr]_m$ and series $\sum_{j=0}^n\Bigl[\E\bigl\{\bGamma_{n+1}\bigr\}{\bigl(\overline\bB^{(1)}\bigr)}^{j}\overline\br^{(1)}\Bigr]_m$ converge for $m = 1,\cdots, NL$. It is known that a sequence is convergent if it is upper bounded and lower bounded by two sequences with the same limit \cite{Sohrab2014}. A series is absolutely convergent if each of its terms is bounded by a term of an absolutely convergent series \cite{Lorenzo2013,Nassif2016Proximal}. Define $s_{2,m}\triangleq\Bigl[\E\bigl\{\bGamma_{n+1}\bigr\}{\bigl(\overline\bB^{(1)}\bigr)}^{j}\overline\br^{(1)}\Bigr]_m$. Since the largest absolute value of the entries in a block vector is smaller than or equal to the block maximum norm of this vector, we have:
\begin{align}
|s_{1,m}|&\!\leq \bigl\|\E\bigl\{\bGamma_{n+1}\bigr\}\bigr\|_{b,\infty}\cdot\bigl\|\overline\bB^{(1)}\bigr\|_{b,\infty}^{n+1}\cdot\bigl\|\E\bigl\{\bv^{(1)}_0\bigr\}\bigr\|_{b,\infty}\notag\\
& \!\overset{\eqref{eq:Bi}}{\leq} \bigl\|\E\bigl\{\bGamma_{n+1}\bigr\}\bigr\|_{b,\infty}\cdot\bigl\|\bM^{(1)}\bigr\|_{b,\infty}^{n+1}\cdot\bigl\|\E\bigl\{\bv^{(1)}_0\bigr\}\bigr\|_{b,\infty}\notag\\
& \!= \! \rho\bigl(\E\bigl\{\bGamma_{n+1}\bigr\}\bigr)\!\cdot\!\bigl[\rho\bigl(\bM^{(1)}\bigr)\bigr]^{n+1}\,\bigl\|\E\bigl\{\bv^{(1)}_0\bigr\}\bigr\|_{b,\infty}\label{eq:s1},\\
|s_{2,m}|& \leq \bigl\|\E\bigl\{\bGamma_{n+1}\bigr\}\bigr\|_{b,\infty}\cdot\bigl\|\overline\bB^{(1)}\bigr\|_{b,\infty}^{j}\cdot\bigl\|\overline\br^{(1)}\bigr\|_{b,\infty}\notag\\
& \overset{\eqref{eq:Bi}}{\leq}\rho\bigl(\E\bigl\{\bGamma_{n+1}\bigr\}\bigr)\cdot \bigl[\rho\bigl(\bM^{(1)}\bigr)\bigr]^{j}\cdot\bigl\|\overline\br^{(1)}\bigr\|_{b,\infty},\label{eq:s2}
\end{align}
where
\vspace{-2mm}
\begin{align}
\bM^{(1)}\triangleq \bI_{NL}-\bU^{(1)}\overline\bH^{(1)}\label{eq:definition-M1},
\end{align}
$\|\cdot\|_{b,\infty}$ is the block maximum norm \cite{Sayed2013intr}, and $\rho(\cdot)$ is the spectral radius of a given matrix. Since $\E\bigl\{\bv^{(1)}_0\bigr\}$ and $\overline\br^{(1)}$ are bounded, it is sufficient to require that $\rho\bigl(\bM^{(1)}\bigr)<1$ and $\rho\bigl(\E\bigl\{\bGamma_{n+1}\bigr\}\bigr)$ is uniformly bounded. It is similar for the second term on the RHS of $\eqref{eq:bvnfinal}$. As discussed later in Section \ref{subsec:combinationlayerAff}, condition \eqref{eq:muConditionAFFPNLMSMean} ensures the convergence of $\E\bigl\{\bGamma_{n+1}\bigr\}$ and $\E\bigl\{\bI_{NL} - \bGamma_{n+1}\bigr\}$. Given that $\nu_{\gamma_k}$ satisfies condition \eqref{eq:muConditionAFFPNLMSMean}, the convergence of \eqref{eq:bvnfinal} requires only $\rho\bigl(\bI_{NL}-\bU^{(i)}\overline\bH^{(i)}\bigr)<1$, which is ensured by step-sizes satisfying \eqref{eq:meanconditionAFF}.

For the asymptotic bias, from \eqref{eq:bvnfinal} and \eqref{eq:compactbvstarexpect2}, we obtain:
\begin{align}
\hspace{-2mm}\E\bigl\{\bv_{\infty}\bigr\} & \!\!=\! \E\bigl\{\bGamma_{\infty}\bigr\}\E\bigl\{\bv_{\infty}^{(1)}\bigr\}  \!+\! (\bI_{NL} \!-\! \E\bigl\{\bGamma_{\infty}\bigr\})\,\E\bigl\{\bv_{\infty}^{(2)}\bigr\}\\
\hspace{-2mm}\E\bigl\{\bv_{\infty}^{(i)}\bigr\} &\!=\! - {\bigl(\bI_{NL}-\overline\bB^{(i)}\bigr)}^{-1}\overline\br^{(i)},
\end{align}
which leads to expression \eqref{eq:biasAFF}.
\vspace{-3mm}
\section*{Appendix D\\Derivation of Equation \eqref{eq:meansquareapprox}}
\label{sec:appendix-DS}

Using \eqref{eq:compactbv}, \eqref{eq:Bf}, \eqref{eq:BG}, and approximation \textbf{Ap}$_4$, we have:
\begin{align}
\E\Bigl\{{\bigl\|\bv_{n+1}^{(i)}\bigr\|}^2_{\bSig^{(i)}_{n+1}}\Bigr\} &= \E\Bigl\{{\bigl\|\bv_{n}^{(i)}\bigr\|}^2_{\bSig^{(i)}_{B,n+1}}\Bigr\}+\trace\Bigl\{\bSig^{(i)}_{n+1}\,\bG^{(i)}\Bigr\}\notag\\
&\quad+
\E\bigl\{\bdf(\br_n^{(i)},\bSig^{(i)}_{n+1},\bv_n^{(i)})\bigr\}\label{eq:meansquarecomicompact}
\end{align}
\vspace{-3mm}with
\begin{align}
\hspace{-3mm}
\bSig^{(i)}_{B,n+1} & \triangleq\E\bigl\{\bB_n^{(i)\top}\bSig^{(i)}_{n+1}\bB_n^{(i)}\bigr\}.
\end{align}
In the derivation of expression \eqref{eq:meansquarecomicompact}, we used relation $\E\{\bg_n^{(i)}\} = 0$, which is derived from \eqref{eq:gin} and \eqref{eq:pzx} directly by using the zero-mean property of $z_{\ell, n}$.

Let $\bsigma_{B,n+1}^{(i)}=\vect\bigl\{\bSig_{B,n+1}^{(i)}\bigr\}$. Under approximation \textbf{Ap}$_4$, and using the properties of $\vect\{\cdot\}$ operator, we have:
\begin{align}
\bsigma_{B,n+1}^{(i)}=\bK^{(i)}\,\bsigma^{(i)}_{n+1},
\end{align}
\vspace{-2mm}where
\begin{align}
\bK^{(i)} & \triangleq\E\bigl\{\bB_n^{(i)\top}\otimes\bB_n^{(i)\top}\bigr\} \approx \overline\bB^{(i)\top}\otimes\overline\bB^{(i)\top}
\end{align}
for sufficiently small step-sizes.
We have:
\begin{equation}
\trace\bigl\{\bSig^{(i)}_{n+1}\,\bG^{(i)}\bigr\}={\bigl[\vect\{\bG^{(i)\top}\} \bigr]}^\top\bsigma^{(i)}_{n+1}.\label{eq:traceSigG}
\end{equation}
For sufficiently small step-sizes, we have:
\begin{align}
\E\bigl\{\bdf(\br_n^{(i)},\bSig^{(i)}_{n+1},\bv_n^{(i)})\bigr\}&\approx\bdf\bigl(\overline\br^{(i)},\bSig^{(i)}_{n+1},\E\bigl\{\bv_n^{(i)}\bigr\}\bigr).\label{eq:Efapprox}
\end{align}
Substituting \eqref{eq:traceSigG} and \eqref{eq:Efapprox} into \eqref{eq:meansquarecomicompact} yields expression \eqref{eq:meansquareapprox}.

\section*{Appendix E\\Proof of Stability Analysis Result 2}
\label{sec:appendix-E}

The convergence of $\E\bigl\{{\|\bv_{n+1}\|}^2_{\bSig}\bigr\}$ requires the convergence of the terms on the RHS of \eqref{eq:meansquareanalysis}. For the last term, using \eqref{eq:meansquareapprox} and iterating from $n=0$, we find that
\begin{align}
&\E\bigl\{\bigl\|\bv_{n+1}^{(1)}\bigr\|^2_{\bsigma^{(1)}_{n+1}}\bigr\} = \bigl\|\bv_{0}^{(1)}\bigr\|^2_{{(\bK^{(1)})}^{n+1}\bsigma^{(1)}_{n+1}}\notag\\
&+{\bigl[\vect\{\bG^{(1)\top}\} \bigr]}^\top\sum_{t=0}^{n}{(\bK^{(1)})}^{t}\bsigma^{(1)}_{n+1}\notag\\&+
\sum_{t=0}^{n}\bdf\bigl(\overline\br^{(1)},{(\bK^{(1)})}^{t}\bsigma^{(1)}_{n+1},\E\bigl\{\bv_{n-t}^{(1)}\bigr\}\bigr)\label{eq:IterationForSquareTerms}
\end{align}
for $n\geq0$ with initial condition $\bv_{0}^{(1)}=\bw_{0}^{(1)}-\bw^{\star}$. Since
\begin{align}
\bsigma^{(1)}_{n+1} =
\E\bigl\{\bGamma_{n+1}^\top\otimes\bGamma_{n+1}^\top\bigr\}\bsigma
\end{align}
where $\bsigma \triangleq \vect\{\bSig\}$, the convergence of \eqref{eq:IterationForSquareTerms} requires that the sequence $s_{3,m}\triangleq\bigl[{(\bK^{(1)})}^{n+1}\,\E\bigl\{\bGamma_{n+1}^\top\otimes\bGamma_{n+1}^\top\bigr\}\bsigma\bigr]_m$ and series $\sum_{t=0}^{n}\bigl[{(\bK^{(1)})}^{t}\E\bigl\{\bGamma_{n+1}^\top\otimes\bGamma_{n+1}^\top\bigr\}\bsigma\bigr]_m$ be convergent for $m=1,\cdots,{(NL)}^2$. Define $s_{4,m}\triangleq\bigl[{(\bK^{(1)})}^{t}\E\bigl\{\bGamma_{n+1}^\top\otimes\bGamma_{n+1}^\top\bigr\}\bsigma\bigr]_m$. Similar to \eqref{eq:s1} and \eqref{eq:s2}, we have:
\begin{align}
|s_{3,m}|& \leq \bigl\|\bK^{(1)}\bigr\|_{b,\infty}^{n+1}\cdot\bigl\|\E\bigl\{\bGamma_{n+1}^\top\otimes\bGamma_{n+1}^\top\bigr\}\bigr\|_{b,\infty}\bigl\|\bsigma\bigr\|_{b,\infty}\notag\\
& = \bigl\|\bK^{(1)}\bigr\|_{b,\infty}^{n+1}\cdot\rho\bigl(\E\bigl\{\bGamma_{n+1}^\top\otimes\bGamma_{n+1}^\top\bigr\}\bigr)\bigl\|\bsigma\bigr\|_{b,\infty}\label{eq:s3},\\
|s_{4,m}|& \leq\bigl\|\bK^{(1)}\bigr\|_{b,\infty}^{t}\cdot\bigl\|\E\bigl\{\bGamma_{n+1}^\top\otimes\bGamma_{n+1}^\top\bigr\}\bigr\|_{b,\infty}\bigl\|\bsigma\bigr\|_{b,\infty}\notag\\
& \!=\!\bigl\|\bK^{(1)}\bigr\|_{b,\infty}^{t}\!\cdot\rho\bigl(\E\bigl\{\bGamma_{n+1}^\top\otimes\bGamma_{n+1}^\top\bigr\}\bigr)\cdot\bigl\|\bsigma\bigr\|_{b,\infty}.\label{eq:s4}
\end{align}
According to assumption \textbf{A2}, we have:
\begin{align}
\E\{\gamma_{k,n+1}\gamma_{\ell,n+1}\}=0,\,\forall k\neq \ell.
\end{align}
Thus, $\rho\bigl(\E\bigl\{\bGamma_{n+1}^\top\otimes\bGamma_{n+1}^\top\bigr\}\bigr)$ is bounded as long as $\E\{{\bGamma_{n+1}\bGamma_{n+1}}\}$ is convergent. We further have:
\begin{align}
\bigl\|\bK^{(1)}\bigr\|_{b,\infty}&\overset{\eqref{eq:ApproxKi}}{=}{\bigl\|\overline\bB^{(1)\top}\otimes\overline\bB^{(1)\top}\bigr\|}_{b,\infty}\notag\\
&\overset{\eqref{eq:Bi}}{\leq} \bigl\|{\pmb{\cal A}}_1^{(1)}\otimes{\pmb{\cal A}}_1^{(1)}\bigr\|_{b,\infty}{\bigl\|\bM^{(1)\top}\otimes\bM^{(1)\top}\bigr\|}_{b,\infty}\notag\\
&\quad\ \ \bigl\|{\pmb{\cal A}}_2^{(1)}\otimes{\pmb{\cal A}}_2^{(1)}\bigr\|_{b,\infty}.
\end{align}
Since $\bigl\|\bsigma\bigr\|_{b,\infty}$ is bounded,
\vspace{-2mm}
\begin{align}
{\bigl\|\bM^{(1)\top}\otimes\bM^{(1)\top}\bigr\|}_{b,\infty}& \overset{\eqref{eq:definition-M1}}{=}\rho\bigl(\bM^{(1)\top}\otimes\bM^{(1)\top}\bigr)\notag\\
& = \bigl[\rho\bigl(\bM^{(1)}\bigr)\bigr]^2
\end{align}
\vspace{-2mm}and
\vspace{-2mm}
\begin{align}
& \bigl\|{\pmb{\cal A}}_j^{(1)}\otimes{\pmb{\cal A}}_j^{(1)}\bigr\|_{b,\infty}  = \bigl\|{\pmb{\cal A}}_j^{(1)}\otimes{\bA}_j^{(1)}\bigr\|_{\infty}\notag\\
& = \max_{\ell,p} \sum_{k = 1}^Na_{j,\ell k}^{(1)}\sum_{q = 1}^Na_{j,pq}^{(1)} = 1,
\end{align}
where the notation $\|\cdot\|_{\infty}$ denotes the maximum absolute row sum of its argument, the convergence of \eqref{eq:s3} and \eqref{eq:s4} requires that $\rho\bigl(\bM^{(1)}\bigr)<1$ and the convergence of $\E\{{\bGamma_{n+1}\bGamma_{n+1}}\}$. It is similar for the first two terms on the RHS of \eqref{eq:meansquareanalysis}. As derived later in Section \ref{subsec:combinationlayerAff}, condition \eqref{eq:muConditionAFFPNLMSMeanaquarestate} guarantees the stabilities of $\E\{{\bGamma_{n+1}}\}$ and $\E\{{\bGamma_{n+1}\bGamma_{n+1}}\}$. Then, for any given weighting matrix $\bSig$, condition \eqref{eq:meanconditionAFF} ensures the mean-square stability of the power-normalized diffusion scheme.
\vspace{-2mm}
\section*{Appendix F\\Recursions for Evaluating Transient MSD}
\label{sec:appendix-F}

Recursion for evaluating $\E\bigl\{\bigl\|\bv_{n}^{(i)}\bigr\|^2_{\bK^{(i)}\bsigma^{(i)}_{n+1}}\bigr\}$ is given in \cite{Chen2015diffusion} as:
\vspace{-3mm}
\begin{align}
\xi_{n+1}^{(i)} &= \xi_{n}^{(i)}+\bigl[{\bigl(\vect\{\bG^{(i)\top}\}\bigr)}^{\top}{(\bK^{(i)})}^{n}\bsigma^{(i)} \notag\\&+ {\|\overline\br^{(i)}\|}^2_{{(\bK^{(i)})}^{n}\bsigma^{(i)}}-{\|\bv_0^{(i)}\|}^2_{(\bI-\bK^{(i)}){(\bK^{(i)})}^{n}\bsigma^{(i)}}\notag\\
&-2\bigl(\bLambda_n^{(i)} + {(\overline\bB^{(i)}\E\{\bv_n^{(i)}\})}^{\top}\otimes\overline\br^{(i)\top}\bigr)\bsigma^{(i)}\bigr],\label{eq:MSDIteration}
\end{align}
and
\begin{align}
\hspace{-3.5mm} \bLambda_{n+1}^{(i)}\!=\!\bLambda_{n}^{(i)}\bK^{(i)}
 \!+\! \bigl({(\overline\bB^{(i)}\E\{\bv_n^{(i)}\})}^{\top}\!\otimes\!\overline\br^{(i)\top}\bigr)\bigl(\bK^{(i)}\!-\!\bI\bigr)
\end{align}
with $\bLambda_{0}^{(i)}=\bO_{1\times{(NL)}^2}$, $\xi_{n+1}^{(i)}=\E\bigl\{\bigl\|\bv_{n+1}^{(i)}\bigr\|^2_{\bK^{(i)}\bsigma^{(i)}_{n+1}}\bigr\}, \xi_{0}^{(i)}=\bigl\|\bv_{0}^{(i)}\bigr\|^2_{\bK^{(i)}\bsigma^{(i)}_{n+1}}, \bsigma^{(i)}=\bK^{(i)}\bsigma^{(i)}_{n+1}$ for $i=1,2$. Besides, \cite{Chen2018Chapter} gives an alternative expression for \eqref{eq:MSDIteration}, which greatly saves computations. Following the same routine, $\E\bigl\{\bv_{n}^{(1)\top}\bSig_{\text{xc},n+1}\bv_{n}^{(2)}\bigr\}$ can be evaluated as follows:
\vspace{-1mm}
\begin{align}
&\xi_{{\rm{x}},n+1} = \xi_{{\rm{x}},n}+\bigl(\vect\{\bG_{\rm x}^{\top}\}\bigr)^{\top}{(\bK_{\rm x})}^{n}\bsigma_{\rm x} +(\bPi_n^{(1)} + \bPi_n^{(2)})\bsigma_{\rm x} \notag\\
&-\bigl[\bigl(\overline\bB^{(2)}\E\{\bv_n^{(2)}\}\bigr)^{\top}\otimes\overline\br^{(1)\top}+\overline\br^{(2)\top}\otimes\bigl(\overline\bB^{(1)}\E\{\bv_n^{(1)}\}\bigr)^{\top}\bigr]\bsigma_{\rm x} \notag\\
&-\bv_0^{(1)\top}\vect^{-1}\bigl\{(\bI-\bK_{\rm x})\bigl({(\bK_{\rm x})}^{n}\bsigma_{\rm x}\bigr)\bigr\}\bv_0^{(2)}\notag\\
&+ \overline\br^{(1)\top}\vect^{-1}\bigl\{{(\bK_{\rm x})}^{n}\bsigma_{\rm x}\bigr\}\overline\br^{(2)},\label{eq:MSDIterationCross}
\end{align}
and
\begin{align}
\bPi_{n+1}^{(1)}&\!=\!\bPi_{n}^{(1)}\bK_{\rm x}\!+\! \bigl[\!\overline\br^{(2)\top}\!\otimes\!(\!\overline\bB^{(1)}\E\{\bv_n^{(1)}\})^{\top}\bigr]\bigl(\!\bI\!-\!\bK_{\rm x}\!\bigr),\\
\bPi_{n+1}^{(2)}&\!=\!\bPi_{n}^{(2)}\bK_{\rm x}\!+\! \bigl[\!(\overline\bB^{(2)}\E\{\bv_n^{(2)}\})^{\top}\!\otimes\!\overline\br^{(1)\top}\bigr]\bigl(\!\bI\!-\!\bK_{\rm x}\!\bigr)\label{eq:MSDIterationCrossPi2}
\end{align}
with $\bPi_{0}^{(i)}=\bO_{1\times{(NL)}^2}$, $\bsigma_{\rm x}=\bK_{\rm x}\bsigma_{{\rm x},n+1}$, $\xi_{{\rm{x}},n+1}=\E\bigl\{\bv_{n+1}^{(1)\top}\bSig_{\text{xc},n+1}\bv_{n+1}^{(2)}\bigr\}, \xi_{{\rm{x}},0}=\E\bigl\{\bv_{0}^{(1)\top}\bSig_{\text{xc},n+1}\bv_{0}^{(2)}\bigr\}$.

\section*{Appendix G\\Proof of Remark 2}
\label{sec:appendix-G}

The steady-state MSD is given by the limit
${\rm MSD}^{\rm steady}=\lim_{n\to\infty}\frac{1}{N}\,\E\bigl\{{\|\bv_{n+1}\|}^2\bigr\}$.
By resorting to expression \eqref{eq:meansquareanalysis}, we obtain:
\vspace{-2mm}
\begin{align}
&\lim_{n\to\infty}\frac{1}{N}\,\E\bigl\{{\|\bv_{n+1}\|}^2\bigr\}= \frac{1}{N}\,\E\Bigl\{{\bigl\|\bv_{\infty}^{(1)}\bigr\|}^2_{\bGamma_{\infty}^\top\bGamma_{\infty}}\Bigr\}\notag\\
&\quad+ \frac{1}{N}\,\E\Bigl\{{\bigl\|\bv_{\infty}^{(2)}\bigr\|}^2_{{(\bI_{NL} - \bGamma_{\infty})}^\top(\bI_{NL} - \bGamma_{\infty})}\Bigr\}\notag\\
&\quad +\frac{2}{N}\,\E\Bigl\{\bv_{\infty}^{(1)\top}\bGamma_{\infty}\,(\bI_{NL} - \bGamma_{\infty})\,\bv_{\infty}^{(2)}\Bigr\},\label{eq:steadyMSD.iteration}
\end{align}
which is the summation of three steady-state values. For the first two terms, recursing \eqref{eq:meansquareapprox} with $n\to\infty$ yields:
\begin{align}
\lim_{n\to\infty}\E&\Bigl\{{\bigl\|\bv_{n+1}^{(i)}\bigr\|}_{(\bI_{(NL)^2}-\bK^{(i)})\bsigma^{(i)}_{n+1}}\Bigr\} = \notag\\ & {\bigl[\vect\{\bG^{(i)\top}\} \bigr]}^\top\bsigma^{(i)}_{\infty}+\bdf\bigl(\overline\br^{(i)},\bSig^{(i)}_{\infty},\E\bigl\{\bv_{\infty}^{(i)}\bigr\}\bigr).\label{eq:steadyMSD.iterationi}
\end{align}
In order to use \eqref{eq:steadyMSD.iterationi} in \eqref{eq:steadyMSD.iteration}, we select $\bsigma^{(1)}_{\infty}$ and $\bsigma^{(2)}_{\infty}$ to satisfy:
\begin{align}
(\bI_{(NL)^2}-&\bK^{(1)})\bsigma^{(1)}_{\infty}=\frac{1}{N}\vect\bigl\{\E\{{\bGamma_{\infty}^\top\bGamma_{\infty}}\}\bigr\},\\
(\bI_{(NL)^2}-&\bK^{(2)})\bsigma^{(2)}_{\infty}=\notag\\&\frac{1}{N}\vect\bigl\{\E\{{(\bI_{NL} - \bGamma_{\infty})}^\top(\bI_{NL} - \bGamma_{\infty})\}\bigr\}.
\end{align}
Similarly for the last term of \eqref{eq:steadyMSD.iteration}, recursing \eqref{eq:crosstermiteration} with $n\to\infty$ yields:
\begin{align}
\hspace{-1mm}&\E\bigl\{\bv_{\infty}^{(1)\top}\vect^{-1}\{(\bI_{(NL)^2}\!-\!\bK_{\text{x}})\bsigma_{\text{x},\infty}\}\bv_{\infty}^{(2)}\bigr\}\!=\!{\bigl[\vect\{\bG_{\text{x}}^\top\}\bigr]}^\top\!\!\bsigma_{\text{x},\infty}\notag\\
&\,+\!\bdf_{\text{x}}\bigl(\overline\br^{(1)},\overline\br^{(2)},\bSig_{\text{x},\infty},\E\bigl\{\!\bv_{\infty}^{(1)}\!\bigr\},\E\bigl\{\!\bv_{\infty}^{(2)}\!\bigr\},\overline\bB^{(1)},\overline\bB^{(2)}\bigr),\label{eq:crossterm-steady}
\end{align}
and we select $\bsigma_{\text{x},\infty}$ to satisfy:
\begin{align}
\hspace{-3mm} (\bI_{(NL)^2}-\bK_{\text{x}})\bsigma_{\text{x},\infty}=
\frac{2}{N} \vect\bigl\{\E\{\bGamma_{\infty}\,(\bI_{NL} - \bGamma_{\infty})\}\bigr\}.
\end{align}
These expressions lead to the MSD at steady state as:
\begin{align}
&{\rm MSD}^{\rm steady}\!=\!\bigl[\vect\{\bG^{(1)\top}\} \bigr]^\top\!\!\bsigma^{(1)}_{\infty}\!+\!\bdf\bigl(\overline\br^{(1)},\bSig^{(1)}_{\infty},\E\bigl\{\bv_{\infty}^{(1)}\bigr\}\bigr)+\notag\\
&\bigl[\!\vect\{\bG^{(2)\top}\!\} \! \bigr]^{\!\top}\!\!\bsigma^{(2)}_{\infty}\!\!+\!\!\bdf\bigl(\!\overline\br^{(2)},\bSig^{(2)}_{\infty},\E\bigl\{\bv_{\infty}^{(2)}\bigr\}\!\bigr)\!\!+\!\!\bigl[\!\vect\{\bG_{\text{x}}^\top\}\!\bigr]^{\!\top}\!\!\bsigma_{{\rm x},\infty} \notag\\
& + \bdf_{\text{x}}\bigl(\overline\br^{(1)},\!\overline\br^{(2)},\!\bSig_{{\rm x},\infty},\E\bigl\{\bv_{\infty}^{(1)}\bigr\},\E\,\bigl\{\bv_{\infty}^{(2)}\bigr\},\!\overline\bB^{(1)},\!\overline\bB^{(2)}\bigr)\label{eq:netSteadyMSD}
\end{align}
\vspace{-2mm}with
\vspace{-2mm}
\begin{align}
&\bSig^{(i)}_{\infty} \!=\! \vect^{-1}\{\bsigma^{(i)}_{\infty}\},\,\,\forall\,\,i = 1, 2\\
&\bSig_{{\rm x},\infty} \!=\! \vect^{-1}\{\bsigma_{{\rm x},\infty}\}.
\end{align}
\vspace{-6mm}
\section*{Appendix H\\Proof of Stability Analysis Results 3 \& 4}
\label{sec:appendix-H}

By substituting \eqref{eq:eakndiffer}, \eqref{eq:equivalentekn} into \eqref{eq:affadapta}, and taking expectation, we obtain:
\begin{align}
	&\E\{\!\gamma_{k,n+1}\!\}\!=\!\E\bigl\{\!\gamma_{k,n}\!\bigr\}\!+\!\E\Bigl\{\!{\nu_{\gamma_k}\over \varepsilon + p_{k,n}}\!\gamma_{k,n}\!\bigl[\widetilde{e}_{k,n}^{(1)}\widetilde{e}_{k,n}^{(2)}\!-\!
(\widetilde{e}_{k,n}^{(1)})^2\bigr]\notag\\& + {\nu_{\gamma_k}\over \varepsilon + p_{k,n}}\,{\bigl(1-\gamma_{k,n}\bigl)}\, \bigl[(\widetilde{e}_{k,n}^{(2)})^2 - \widetilde{e}_{k,n}^{(1)}\widetilde{e}_{k,n}^{(2)}	
\bigr]\Bigr\}.\label{eq:IterTheoreticalAff}
\end{align}
By resorting to EMSEs $J_{\text{ex},k,n}^{(i)}$ and cross-EMSE $J_{\text{ex},k,n}^{(1,2)}$ defined in Section \ref{subsec:affuniversality}, as well as \textbf{Ap}$_2$, \textbf{Ap}$_5$, \textbf{Ap}$_6$, expression \eqref{eq:IterTheoreticalAff} becomes
\begin{align}
	\E\{\gamma_{k,n+1}\}
&=\E\bigl\{\gamma_{k,n}\bigr\}\Bigl[1 - {\nu_{\gamma_k}\over \varepsilon + \bar p_{k,n}}\,\bigl({\triangle J_{k,n}^{(1)}+\triangle J_{k,n}^{(2)}}\bigr)\Bigl] \notag\\&\quad+ {\nu_{\gamma_k}\over \varepsilon + \bar p_{k,n}}\,\bigl(J_{\text{ex},k,n}^{(2)} - J_{\text{ex},k,n}^{(1,2)}\bigr),\label{eq:meanIterationAFF}
\end{align}
where we adopt the following approximation to simplify the derivation:
\begin{align}
&\E\Bigl\{{\nu_{\gamma_k}\over \varepsilon + p_{k,n}}\Bigr\} \approx {\nu_{\gamma_k}\over \varepsilon + \bar p_{k,n}}\quad{\rm with}\notag\\&\bar p_{k,n} = \eta\,\bar p_{k,n-1} + (1 - \eta)({\triangle J_{k,n}^{(1)}+\triangle J_{k,n}^{(2)}}).\label{eq:approximationStep}
\end{align}
From~\eqref{eq:meanIterationAFF} a sufficient condition can be derived for the step-size $\nu_{\gamma_k}$ to ensure the mean stability of the power-normalized scheme \cite{Candido2010,Sayed2008}:
\vspace{-1.5mm}
\begin{align}
\Bigl|1 - {\nu_{\gamma_k}\over \varepsilon + \bar p_{k,n}}\,\bigl({\triangle J_{k,n}^{(1)}+\triangle J_{k,n}^{(2)}}\bigr)\Bigl|<1 - \phi,\,\,\forall n,\label{eq:ConverRadiusPN}
\end{align}
with a positive constant $\phi$, i.e., the left term is uniformly bounded away from one. A sufficient condition to ensure \eqref{eq:ConverRadiusPN} is given by:
\vspace{-2.5mm}
\begin{align}
0<\nu_{\gamma_k}<\min_n\biggl\{\frac{\varepsilon + \bar p_{k,n}}{\triangle J_{k,n}^{(1)}+\triangle J_{k,n}^{(2)}}\biggr\}.\label{eq:SuffiConditionPN}
\end{align}
Substituting $\bar p_{k,n}$ of \eqref{eq:approximationStep} into \eqref{eq:SuffiConditionPN}, and after some mathematical manipulations, we obtain condition \eqref{eq:muConditionAFFPNLMSMean}.  By further taking the limit of~\eqref{eq:meanIterationAFF} as $n\to\infty$, and solving for $\E\{\gamma_{k,\infty}\}$, we arrive at equation~\eqref{eq:affStationarySolution}.

Next, we evaluate the mean-square behavior of $\gamma_{k,n}$. To simplify the notation, define $\nu_{\gamma_{k,n}}\triangleq {\nu_{\gamma_k}\over \varepsilon + p_{k,n}}$. By substituting \eqref{eq:eakndiffer} and \eqref{eq:equivalentekn} into \eqref{eq:affadapta}, we obtain:
\begin{align}
\gamma_{k,n+1}
&=\underbrace{\gamma_{k,n}\Bigl[1-{\nu_{\gamma_{k,n}}}\,\bigl(\widetilde{e}_{k,n}^{(2)} - \widetilde{e}_{k,n}^{(1)}\bigr)^2\Bigr]}_{\alpha_1} \notag\\&\quad+ \underbrace{{\nu_{\gamma_{k,n}}}\,\Bigl[{(\widetilde{e}_{k,n}^{(2)})}^2-\widetilde{e}_{k,n}^{(1)}\widetilde{e}_{k,n}^{(2)}\Bigr]}_{\alpha_2}\notag\\&\quad + \underbrace{{\nu_{\gamma_{k,n}}}\,\bigl(\widetilde{e}_{k,n}^{(2)} - \widetilde{e}_{k,n}^{(1)}\bigr)z_{k,n}}_{\alpha_3}.\label{eq:gammakn}
\end{align}
As a consequence, we have:
\begin{align}
\hspace{-3mm} \E\{\gamma_{k,n+1}^2\} \!=\!
 \E\bigl\{\alpha_1^2\bigr\} \!+\! \E\bigl\{\alpha_2^2\bigr\} \!+\! \E\bigl\{\alpha_3^2\bigr\} \!+\! 2\E\bigl\{\alpha_1\alpha_2\bigr\}.\label{eq:AFFPNLMSExpectation}
\end{align}
By resorting to approximations \textbf{Ap}$_2$, \textbf{Ap}$_5$--\textbf{Ap}$_7$, and using expression \eqref{eq:gammakn}, we obtain:
\begin{align}
\E\bigl\{\alpha_1^2\bigr\}
&= \E\{\gamma_{k,n}^2\}\Bigl[1 + 3\E\bigl\{\nu^2_{\gamma_{k,n}}\bigr\} \bigl({\triangle J_{k,n}^{(1)}+\triangle J_{k,n}^{(2)}}\bigr)^2\notag\\& \quad - 2\E\bigl\{{\nu_{\gamma_{k,n}}}\bigr\}\,{\bigl({\triangle J_{k,n}^{(1)}+\triangle J_{k,n}^{(2)}}\bigr)}\Bigr],\label{eq:D2}\\
\E\bigl\{\alpha_2^2\bigr\}
&= \E\bigl\{\nu^2_{\gamma_{k,n}}\bigr\}\,J_{\text{ex},k,n}^{(2)}\,\bigl({\triangle J_{k,n}^{(1)}+\triangle J_{k,n}^{(2)}}\bigr)\notag\\
& \quad + 2\E\bigl\{\nu^2_{\gamma_{k,n}}\bigr\}\,{\bigl(\triangle J_{k,n}^{(2)}\bigr)}^2,\label{eq:E2}\\
\E\bigl\{\alpha_3^2\bigr\}
&= \sigma_{z,k}^2\E\bigl\{\nu^2_{\gamma_{k,n}}\bigr\}\bigl({\triangle J_{k,n}^{(1)}+\triangle J_{k,n}^{(2)}}\bigr),\label{eq:F2}\\
\E\bigl\{\alpha_1\alpha_2\bigr\}
& = \E\bigl\{\gamma_{k,n}\bigr\}\Bigl[\E\bigl\{{\nu_{\gamma_{k,n}}}\bigr\}\,\triangle J_{k,n}^{(2)} \notag\\
&\quad - 3\E\bigl\{\nu^2_{\gamma_{k,n}}\bigr\} {\bigl({\triangle J_{k,n}^{(1)}+\triangle J_{k,n}^{(2)}}\bigr)}\triangle J_{k,n}^{(2)}\Bigr].\label{eq:DE}
\end{align}
By adopting approximation
$\E\bigl\{\nu^2_{\gamma_{k,n}}\bigr\}\approx {\bigl[\E\bigl\{{\nu_{\gamma_{k,n}}}\bigr\}\bigr]}^2$
in \eqref{eq:D2}--\eqref{eq:DE}, we obtain the expression of $\E\{\gamma_{k,n+1}^2\}$.  Taking the limit of \eqref{eq:AFFPNLMSExpectation} with $n\to\infty$, and solving for $\E\{\gamma_{k,\infty}^2\}\triangleq\lim_{n\to\infty}\E\{\gamma_{k,n}^2\}$, we obtain the steady-state value \eqref{eq:gammasquaresteadyAff}, with $\bar\nu_{\infty}\triangleq \E\{\nu_{\gamma_{k,\infty}}\} = \lim_{n\to\infty}\E\{\nu_{\gamma_{k,n}}\}$.

From \eqref{eq:AFFPNLMSExpectation} and the term $\E\bigl\{\alpha_1^2\bigr\}$, the range of step-sizes to ensure the mean-square stability is given by \eqref{eq:conditionMeanSquare}.
\begin{figure*}[!t]
\normalsize
\begin{equation}
\Bigl|1 + 3{\Bigl({\nu_{\gamma_k}\over \varepsilon + \bar p_{k,n}}\Bigr)}^2\,{\Bigl({\triangle J_{k,n}^{(1)}+\triangle J_{k,n}^{(2)}}\Bigr)}^2- {2\nu_{\gamma_k}\over \varepsilon + \bar p_{k,n}}\,\bigl({\triangle J_{k,n}^{(1)}+\triangle J_{k,n}^{(2)}}\bigr)\Bigl|<1 - \phi,\,\,\forall n\label{eq:conditionMeanSquare}
\end{equation}
\hrulefill
\vspace{-4mm}
\end{figure*}
With the use of approximation \eqref{eq:approximationStep}, then for a constant step-size, a sufficient condition to ensure \eqref{eq:ConverRadiusPN} is given by:
\begin{align}
0<\nu_{\gamma_k}<\min_n\biggl\{\frac{\varepsilon + \bar p_{k,n}}{3\bigl(\triangle J_{k,n}^{(1)}+\triangle J_{k,n}^{(2)}\bigr)}\biggr\}\label{eq:SufficonditionMeanSquare}.
\end{align}
Substituting $\bar p_{k,n}$ of \eqref{eq:approximationStep} into \eqref{eq:SufficonditionMeanSquare}, and after some mathematical manipulations, we arrive at condition \eqref{eq:muConditionAFFPNLMSMeanaquarestate}.
\vspace{-3mm}
\section*{Appendix I\\Proof of Stability Analysis Results 7 \& 8}
\label{sec:appendix-I}

We first evaluate the mean behavior of $\gamma_{k,n}$ for the sign-regressor diffusion scheme in \eqref{eq:adaptExpAffSR}.
To make the analysis tractable, we introduce approximation \textbf{Ap}$_5$ and joint Gaussian assumption \textbf{Ap}$_7$ in the same way. Furthermore, by resorting to Price's theorem \cite{Sayed2008}, the following approximations hold:
\vspace{-2mm}
\begin{align}
&\E\Bigl\{\!\widetilde{e}_{k,n}^{(2)}\!\sgn\bigl\{\widetilde{e}_{k,n}^{(2)} \!-\! \widetilde{e}_{k,n}^{(1)}\bigr\}\!\Bigr\}\!\!\approx \!\!\sqrt{\frac{2}{\pi}}\frac{\triangle J_{k,n}^{(2)}}{\sqrt{{\triangle J_{k,n}^{(1)}\!+\!\triangle J_{k,n}^{(2)}}}}\label{eq:approximationresult1}\\
&\E\bigl\{\bigl|\widetilde{e}_{k,n}^{(2)} - \widetilde{e}_{k,n}^{(1)}\bigr|\bigr\}\approx\sqrt{{2\,\bigl({\triangle J_{k,n}^{(1)}+\triangle J_{k,n}^{(2)}}\bigr)}/{\pi}}.\label{eq:approximationresult2}
\end{align}
It is noted that approximation \eqref{eq:approximationresult1} is used in evaluating the transient and steady-state behaviors of $\E\bigl\{\gamma_{k,n}\bigr\}$ and $\E\{\gamma_{k,n}^2\}$.
Then, under \textbf{Ap}$_5$ and \textbf{Ap}$_7$, and with the use of approximation \eqref{eq:approximationresult2}, iteration \eqref{eq:adaptExpAffSR} becomes:
\vspace{-2mm}
\begin{align}
\E\{\gamma_{k,n+1}\}
&\approx\E\bigl\{\gamma_{k,n}\bigr\} + \nu_{\gamma_{k}}\E\Bigl\{\widetilde{e}_{k,n}^{(2)}\sgn\bigl\{\widetilde{e}_{k,n}^{(2)} \!-\! \widetilde{e}_{k,n}^{(1)}\bigr\}\Bigr\}\notag\\
&-{\nu_{\gamma_k}}\E\bigl\{\gamma_{k,n}\bigr\}\sqrt{{2\bigl({\triangle J_{k,n}^{(1)}+\triangle J_{k,n}^{(2)}}\bigr)}/{\pi}}.\label{eq:IterTheoreticalAffSRresult}
\end{align}
To ensure the mean stability of $\gamma_{k,n}$, the step-size parameter ${\nu_{\gamma_k}}$ must satisfy:
\vspace{-2mm}
\begin{align}
\Bigl|1-{\nu_{\gamma_k}}\,\sqrt{{2\,\bigl({\triangle J_{k,n}^{(1)}+\triangle J_{k,n}^{(2)}}\bigr)}/{\pi}}\,\Bigr|<1-  \phi,\,\,\forall n \label{eq:AffSRstepsize}
\end{align}
with a positive $\phi$. A sufficient condition is then given by:
\vspace{-2mm}
\begin{align}
0<{\nu_{\gamma_k}}<\frac{1}{\max_{n} \Bigl\{\sqrt{{2\,\bigl({\triangle J_{k,n}^{(1)}+\triangle J_{k,n}^{(2)}}\bigr)}/{\pi}}\Bigr\}},
\end{align}
which leads to condition \eqref{eq:AffSRstepsizeConditionMean} directly.
In addition, by taking the limit of \eqref{eq:IterTheoreticalAffSRresult} as $n\to\infty$, we obtain the same steady-state value for $\E\{\gamma_{k,n}\}$ as in \eqref{eq:affStationarySolution}.

To evaluate the mean-square behavior of $\gamma_{k,n}$, by substituting \eqref{eq:eakndiffer}, \eqref{eq:equivalentekn} into \eqref{eq:adaptaAFFSRLMS} and rearranging terms, we obtain:
\begin{align}
\gamma_{k,n+1}
& =\underbrace{\bigl(1-{\nu_{\gamma_k}}\,\bigl|\widetilde{e}_{k,n}^{(2)} - \widetilde{e}_{k,n}^{(1)}\bigr|\bigr)\,\gamma_{k,n}}_{\alpha_4}\notag\\
&\quad+ \underbrace{{\nu_{\gamma_k}} \widetilde{e}_{k,n}^{(2)}\sgn\bigl\{\widetilde{e}_{k,n}^{(2)} - \widetilde{e}_{k,n}^{(1)}\bigr\}}_{\alpha_5}\notag\\
&\quad+ \underbrace{{\nu_{\gamma_k}}z_{k,n}\sgn\bigl\{\widetilde{e}_{k,n}^{(2)} - \widetilde{e}_{k,n}^{(1)}\bigr\}}_{\alpha_6}.\label{eq:adaptAffSRmeansquare}
\end{align}
By squaring \eqref{eq:adaptAffSRmeansquare} and taking expectation, we have
\begin{align}
\E\{\gamma_{k,n+1}^2\}
\!=\!\E\{\alpha_4^2\} + \E\{\alpha_5^2\}
  + \E\{\alpha_6^2\} + 2\E\{\alpha_4\alpha_5\}.\label{eq:adaptAffSRmeansquareexpect}
\end{align}
To obtain an explicit expression for the mean-square behavior of $\gamma_{k,n}$, we have to evaluate the terms on RHS of \eqref{eq:adaptAffSRmeansquareexpect} one by one under the joint Gaussian assumption \textbf{Ap}$_7$. By resorting to \textbf{Ap}$_5$, \textbf{Ap}$_7$ and utilizing \eqref{eq:adaptAffSRmeansquare}, we obtain:
\vspace{-2mm}
\begin{align}
\E\{\!\alpha_4\alpha_5\!\}\!&\!=\!\E\{\!\gamma_{k,n}\!\}\!\Bigl[\nu_{\gamma_{k}}\E\bigl\{\widetilde{e}_{k,n}^{(2)}\sgn\{\!\widetilde{e}_{k,n}^{(2)} \!\!-\! \widetilde{e}_{k,n}^{(1)}\!\}\!\bigr\}\!\!-\!\!\nu_{\gamma_{k}}^2\!{\triangle\!J_{k,n}^{(2)}}\Bigr] \notag\\
&\!\!\!\!\approx\!\E\{\gamma_{k,n}\}\!\biggl[\!\sqrt{\frac{2}{\pi}}\frac{\nu_{\gamma_k}\,{\triangle J_{k,n}^{(2)}}}{\!\!\!\sqrt{{\triangle J_{k,n}^{(1)}\!\!+\!\triangle J_{k,n}^{(2)}}}}\!-\!\nu_{\gamma_k}^2{\triangle J_{k,n}^{(2)}}\!\biggr],\label{eq:GHcrossterm}\\
\E\{\alpha_4^2\}
&\approx\E\{\gamma_{k,n}^2\}\Bigl[1+\nu_{\gamma_k}^2\bigl({\triangle J_{k,n}^{(1)}+\triangle J_{k,n}^{(2)}}\bigr)\notag\\
&\quad-2{\nu_{\gamma_k}}\,\sqrt{{2\,\bigl({\triangle J_{k,n}^{(1)}+\triangle J_{k,n}^{(2)}}\bigr)}/{\pi}}\,\Bigr],\label{eq:GSquare}\\
\E\{\alpha_5^2\}&\approx \nu_{\gamma_k}^2J_{\text{ex},k,n}^{(2)},\label{eq:HSquare}\\
\E\{\alpha_6^2\}&\approx \nu_{\gamma_k}^2\sigma_{z,k}^2,\label{eq:ISquare}
\vspace{-2mm}
\end{align}
and we used ${\bigl[\sgn\bigl\{\widetilde{e}_{k,n}^{(2)} - \widetilde{e}_{k,n}^{(1)}\bigr\}\bigr]}^2\approx 1$ in the derivations of \eqref{eq:HSquare} and \eqref{eq:ISquare}. Equations \eqref{eq:adaptAffSRmeansquareexpect} and \eqref{eq:GHcrossterm}--\eqref{eq:ISquare} constitute the iteration of the mean-square behavior of $\gamma_{k,n}$. From \eqref{eq:adaptAffSRmeansquareexpect}  and the term $\E\{\alpha_4^2\}$, we obtain the condition to ensure the mean-square stability of $\gamma_{k,n}$ as:
\vspace{-1mm}
\begin{equation}
\Bigl|1+\nu_{\gamma_k}^2\bigl({\triangle J_{k,n}^{(1)}\!+\!\triangle J_{k,n}^{(2)}}\bigr)-{\nu_{\gamma_k}}\!\sqrt{{8\bigl({\triangle J_{k,n}^{(1)}\!+\!\triangle J_{k,n}^{(2)}}\bigr)}/{\pi}}\Bigr|\!\!<\!\!1- \phi\label{eq:AffSRmeansquare}
\end{equation}
for all $n$. For a constant step-size, a sufficient condition to ensure \eqref{eq:AffSRmeansquare} is given by equation \eqref{eq:AffSRmeansquareConditionMeanaquarestate}.
In addition, by taking the limit of \eqref{eq:adaptAffSRmeansquareexpect} as $n\to\infty$, we obtain the steady-state value of $\E\{\gamma_{k,n}^2\}$ as \eqref{eq:AffSRmeansquareSteadyState}.

\bibliographystyle{IEEEbib}
\bibliography{reference}

\end{document}